\documentstyle[12pt]{article}

\begin{document}
\input{psfig.sty}
\begin{flushright}
\baselineskip=12pt
UPR-957-T \\
\end{flushright}

\begin{center}
\vglue 1.5cm
{\Large\bf Discrete Symmetry and GUT Breaking}
\vglue 2.0cm
{\Large Tianjun Li~\footnote{E-mail: tli@bokchoy.hep.upenn.edu,
phone: (215) 573-5820, fax: (215) 898-2010.}}
\vglue 1cm
{ Department of Physics and Astronomy \\
University of Pennsylvania, Philadelphia, PA 19104-6396 \\  
U.  S.  A.}
\end{center}

\vglue 1.5cm
\begin{abstract}
We study the supersymmetric GUT models where the
supersymmetry and GUT gauge symmetry can be broken by the
discrete symmetry. First,
with the ansatz that there exist discrete symmetries in the 
 branes' neighborhoods, we discuss the general
 reflection $Z_2$ symmetries and
GUT breaking on $M^4\times M^1$ and 
$M^4\times M^1\times M^1$. In those models, 
the extra dimensions can be large and
the KK states can be set arbitrarily heavy. Second, considering
the extra space manifold is the annulus $A^2$ or disc $D^2$, we can 
define any $Z_n$ symmetry
and break any 6-dimensional $N=2$
 supersymmetric $SU(M)$ models down to the 4-dimensional $N=1$
supersymmetric $SU(3)\times SU(2)\times U(1)^{M-4}$ models for the zero
modes.
In particular, there might exist the interesting 
scenario on $M^4\times A^2$ where just
a few KK states are light, while the others are relatively heavy.
Third, we discuss the complete global discrete symmetries on 
$M^4\times T^2$ and study the GUT breaking.
\\[1ex]
\\[1ex]

\end{abstract}

\vspace{0.5cm}
\begin{flushleft}
\baselineskip=12pt
October 2001\\
\end{flushleft}
\newpage
\setcounter{page}{1}
\pagestyle{plain}
\baselineskip=14pt

\section{Introduction}
Grand Unified Theory (GUT) gives us an simple
 and elegant understanding of the quantum numbers of the quarks and
leptons,
and the success of gauge coupling unification in the Minimal
Supersymmetric
Standard Model strongly support
 this idea. The Grand Unified Theory at high energy scale has
been widely accepted now, there are some problems in GUT: 
 the grand unified
gauge symmetry breaking mechanism, the doublet-triplet splitting problem,
and
the proton decay, etc.

As we know, one obvious approach to break GUT gauge symmetry is 
Higgs mechanism~\cite{HIGGSPI}, which is discussed extensively in
 phenomenology. Another
approach is spin connection embedding, which is used in the 
weakly coupled heterotic string $E_8\times E_8$, and M-theory on 
$S^1/Z_2$~\cite{ESIX}. Because the Calabi-Yau manifold has $SU(3)$
holonomy,
the observable $E_8$ gauge group can be broken down to 
$E_6$ by spin connection embedding. In addition, the GUT gauge symmetry
can be broken down to a low energy subgroup by means of Wilson lines,
provided that the fundamental group of the extra space manifold or
orbifold is non-trivial~\cite{WLSB}.

Recently, a new scenario to explain above questions in GUT has
 been suggested 
by Kawamura~\cite{kaw1, kaw2, kaw3},
 and further discussed by a lot of papers~\cite{SBPL, LIT2}. 
 The key point is that the GUT
gauge symmetry exists in 5 or higher dimensions and is broken down to the
4-dimensional 
$N=1$ supersymmetric Standard Model like gauge symmetry for
 the zero modes due to the
 discrete symmetries in the branes' neighborhoods, which
become the non-trivial orbifold
 projections on the multiplets and gauge generators in GUT. 
So, we would like to call it the discrete symmetry approach.
The attractive models have been constructed explicitly, where
the supersymmetric 5-dimensional and 6-dimensional 
GUT models are broken down to
the 4-dimensional $N=1$
 supersymmetric $SU(3)\times SU(2) \times U(1)^{n-3}$
model, where $n$ is the rank of GUT group, through the 
compactification on various orbifolds. 
The GUT gauge symmetry breaking and doublet-triplet
splitting problems have been solved neatly by the orbifold projections,
and other interesting phenomenology, like $\mu$ problems, gauge coupling
unifications, non-supersymmetric GUT, gauge-Higgs unification,
proton decay, etc, have also been discussed~\cite{SBPL, LIT2}. 
By the way, it seems to us that this approach is similar to
the Wilson line approach, but not the same, for example,
the fundamental group of extra space manifold can be 
trivial in the discrete symmetry approach, and
we may not break the supersymmetry by Wilson line approach.
 
On the other hand, large extra dimension scenarios with 
branes have been an very interesting subject
for the past few years, where the gauge hierarchy problem 
can be solved because the 
physical volume of extra dimensions may be very large and 
the higher dimensional Planck scale might be low~\cite{AADD}, 
or the metric for
the extra dimensions has warp factor~\cite{LRRS}. Naively, 
one might think the 
masses of KK states are $\sqrt {\sum_i n_i^2/R_i^2}$, where 
$R_i$ is the radius of the
$i-th$ extra dimension. However, it is shown that this is not 
true if one considered
the shape moduli~\cite{KRD} or the local discrete symmetry in the
brane neighborhood~\cite{LIBLS}, and it may be possible to maintain the 
ratio (hierarchy) between the higher
dimensional Planck scale and 4-dimensional Planck scale 
while simultaneously 
making the KK states arbitrarily heavy.
So, a lot of experimental bounds on the
theories with large extra dimensions are relaxed.
Moreover, the gauge symmetry and supersymmetry can be broken if
we consider the local discrete symmetry~\cite{LIBLS}.

In this paper, we study the supersymmetric GUT models where the
supersymmetry and GUT gauge symmetry can be broken by the
discrete symmetries in the branes' neighborhoods
or on the extra space manifold. We require that for the zero modes
in the bulk, the supersymmetric 
GUT models are broken down to the 4-dimensional $N=1$
 supersymmetric $SU(3)\times SU(2) \times U(1)^{n-3}$ model, and
above the GUT scale or including the zero modes and KK modes,
 the bulk should preserve the original GUT gauge symmetry
and supersymmetries, i. e., we can not project out all the zero modes
and KK modes of the fields in the theories.
In addition, we define two discrete symmetries $Z_n$ and $Z_n'$
are equivalent if: 
(1) $n=n'$; (2) in order to satisfy our requirement, the
representation for the generator of $Z_n$ in the adjoint representation
of GUT group $G$ must be the same as that for the
generator of $Z_n'$.

First, we would like to explore the general scenarios
where the GUT gauge symmetry and supersymmetry can be broken 
by the  discrete symmetries 
in the brane neighborhood, and the masses of KK states can be 
set arbitrarily heavy. 
Our ansatz is that there exist the discrete symmetries (local or global)
in the special branes' neighborhoods,
which become the additional constraints on the KK states. 
 The KK states, which satisfy the discrete
symmetries, remain in the theories, while the KK states, which do
not satisfy the discrete symmetries, are projected out.
Therefore, we can construct the theories with only zero modes for all the
KK modes are projected out, or the theories which have large extra
dimensions
and arbitrarily heavy KK modes because there is no simple relation between
the
mass scales of extra dimensions and the masses of KK states.
In addition, the bulk gauge symmetry and supersymmetry can be broken on
the
special branes for the zero and KK modes, and in the bulk for the zero 
modes by local and global discrete symmetries.

(I) We generalize our previous models~\cite{LIBLS} to the
models on the space-time $M^4\times S^1$
and $M^4\times I^1$, where the $M^4$ is the 4-dimensional Minkowski 
space-time. And we point out that the general models on 
$M^4\times I^1$ can be obtained from the general models on
$M^4\times S^1$ by moduloing the $(Z_2)^{k_0}$ symmetry
in which $k_0$ is the positive integer. Moreover,
we find that, to satisfy our ansatz and requirement,
there are at most two non-equivalent $Z_2$ symmetries, which can
be local or global. Therefore,
we can only discuss the 5-dimensional $N=1$ supersymmetric $SU(5)$ model. 
In this scenario, the bulk 4-dimensional $N=2$ supersymmetry and 
$SU(5)$ gauge symmetry are broken down to the 4-dimensional $N=1$ 
supersymmetry and
$SU(3)\times SU(2)\times U(1)$ gauge symmetry on 
the special brane with GUT breaking $Z_2$ symmetry for all the modes, 
and in the bulk for the zero modes. And the 3-branes preserve
half of the bulk supersymmetry.
 Moreover, the masses of KK states can be set 
arbitrarily heavy although the physical size
of the fifth dimension can be large, even at millimeter range.
By the way, 
one can also discuss the non-supersymmetric $SU(6)$ and $SO(10)$ breaking,
 however, there are zero modes for $A_5^{\hat a}$ where ${\hat a}$ is the
index
 related to the broken gauge generators under two 
non-equivalent $Z_2$ projections.

(II) We study the models on the space-time $M^4\times M^1\times M^1$
where $M^1$ can be $S^1$, $S^1/Z_2$, and $I^1$. Because the
extra space manifold is the product of two 1-dimensional manifold
and the discussions are similar, 
as representatives, we  discuss the models
on the space-times $M^4\times S^1\times S^1$
where there are parallel 4-branes with $Z_2$ symmetry 
 along the fifth and sixth dimensions, and there
are at most four non-equivalent $Z_2$ symmetries. We also discuss the
models on the space-time $M^4\times S^1/Z_2\times S^1/Z_2$ where
there are only four 4-branes at boundaries and 
some 3-branes in the bulk, and there are three non-equivalent $Z_2$
symmetries. 
In those models, the extra dimensions can be 
large and the masses of KK states can be set arbitrarily heavy.
The 6-dimensional non-supersymmetric GUT models and $N=1$ 
supersymmetric GUT models can be considered as special cases of
$N=2$ supersymmetric GUT models, 
 so, we discuss the 6-dimensional $N=2$ supersymmetric GUT models.
Because $N=2$ 6-dimensional supersymmetric theory has 16 real
supercharges,
which corresponds to $N=4$ 4-dimensional supersymmetric theory,
we can not have hypermultiplets in the bulk, and then,
we have to put the Standard Model fermions on the brane.
As representatives, we discuss the 6-dimensional $N=2$ supersymmetric
$SU(6)$ and $SO(10)$ models on the space-time $M^4\times S^1\times S^1$,
and the 6-dimensional $N=2$ supersymmetric $SU(6)$ models with gauge-Higgs
unification on the space-time $M^4\times S^1/Z_2\times S^1/Z_2$.
For the zero modes, the bulk 4-dimensional $N=4$  supersymmetry and
$SU(6)$ or $SO(10)$ gauge symmetry are broken down to the
4-dimensional $N=1$ supersymmetry and 
$SU(3)\times SU(2) \times U(1)^2$ gauge symmetry.

Second, we discuss the models where the extra space
manifold is the disc $D^2$ or annulus $A^2$. In this 
kind of scenarios, we
can naturally use the complex coordinates and introduce global
$Z_n$ symmetry for any positive integer
$n$, so, we can break any $SU(M)$ gauge symmetry for $M~\ge~5$
 down to the $SU(3)\times SU(2)\times U(1)^{M-4}$ gauge symmetry.
Similar to above, we only study the
 6-dimensional $N=2$ supersymmetric GUT models
with the Standard Model fermions on the boundary 4-branes
or on the 3-brane at origin if the extra space manifold is disc $D^2$.
 There are 4-dimensional $N=1$ supersymmetry and
$SU(3)\times SU(2) \times U(1)^{M-4}$ gauge symmetry in the bulk and on
the 4-branes for the zero modes, and on the 3-brane at origin in 
the disc $D^2$ scenario.
Including all the KK states,  
we will have 4-dimensional $N=4$ supersymmetry and
 $SU(M)$ gauge symmetry in the bulk, and on
the 4-branes. By the way,
 if we put the Standard Model fermions on the 3-brane at origin,
 the extra dimensions
can be large and the gauge hierarchy problem can be solved 
for there does not exist the proton decay problem at all.
As an example, we discuss the 
6-dimensional $N=2$ supersymmetric $SU(6)$ model
on $M^4\times A^2$ or $M^4\times D^2$ with $Z_9$
symmetry. Moreover, if the extra space manifold is annulus $A^2$,
 for suitable choices of the inner radius and outer radius,
we might construct the models where only a few KK states are light,
and the other KK states are relatively heavy due to the boundary
condition on the inner and outer boundaries, so, we might
produce the light KK states of gauge fields at future
colliders, which is very interesting in collider physics.

In addition, if the extra space manifold is 
an sector of $D^2$ or an segment of $A^2$, we point out that
 the masses of KK states can be set arbitrarily heavy
if the range of angle is small enough.
However, we can not define the discrete symmetry $Z_n$ for $n > 2$
on the sector of $D^2$ or  segment of $A^2$,
 so, it is not interesting for us to discuss the
supersymmetric GUT breaking in this case.

Third, we discuss the complet global discrete symmetry 
on the space-time $M^4\times T^2$.
We prove that the possible global discrete symmetries on the torus is
$Z_2$,
$Z_3$, $Z_4$, and $Z_6$. We also discuss the 6-dimensional $N=2$
supersymmetric $SU(5)$ models on the space-time
$M^4\times T^2$ with $Z_6$ symmetry, where the Standard Model
fermions on the obsevable 3-brane at one of the fixed points.
There are 4-dimensional $N=1$ supersymmetry and
the Standard Model gauge symmetry in the bulk for the
zero modes, and on the 3-brane
at $Z_6$ fixed point for all the modes. Including the
KK states, we will have the 4-dimensional $N=4$ supersymmetry and
 $SU(5)$ gauge symmetry in the bulk, the
4-dimensional $N=1$ supersymmetry and $SU(5)$ gauge symmetry on
the 3-branes at $Z_3$ fixed points, and the 4-dimensional
$N=4$ supersymmetry and $SU(3)\times SU(2) \times U(1)$ gauge 
symmetry on the 3-branes at $Z_2$ fixed points.
The Standard Model fermions and Higgs fields can be on any 3-brane
at one of the fixed points. In particular, if we put the Standard Model
fermions and Higgs fields on the 3-brane at $Z_6$ fixed point, the
extra dimensions can be large and the gauge hierarchy problem can be
solved
because there is no proton decay problem at all.

This paper is organized as follows: in section 2,  
we discuss the discrete symmetry in the brane neighborhood in general.
We study the discrete symmetry on the space-time $M^4\times M^1$,
$M^4\times M^1\times M^1$, $M^4\times A^2$, $M^4\times T^2$ in
sections 3, 5, 7, 9, respectively. And we discuss the supersymmetric
GUT breaking on the space-time $M^4\times M^1$,
$M^4\times M^1\times M^1$, $M^4\times A^2$, $M^4\times T^2$ in sections 4,
6, 8, 10
respectively.
 Our discussion and conlcusion are given in section 11.

\section{Discrete Symmetry in the Brane Neighborhood}
We assume that in a (4+n)-dimensional space-time manifold $M^4\times M^n$
where 
$M^4$ is the 4-dimensional Minkowski space-time and $M^n$ is the manifold
for extra space dimensions, there exist some topological defects, or
we call them branes for simplicity.
The special branes, which we are interested in, have 
co-dimension one or more than
one. Assuming we have $K$ special branes and using the $I-th$ 
special brane as a representative,
 our ansatz is that in the open neighborhood
 $M^4\times U_I$ ($U_I$ $\subset $ $M^n$)
of the $I-th$ special brane, there is a global or local discete 
symmetry\footnote{Global discrete symmetry is a
``special'' case of local discrete symmetry. The key difference is that,
the 
space-time manifold can modulo the global discrete symmetry and become a
quotient space-time manifold or orbifold.}, which forms a discrete
group $\Gamma_I$, where $I=1, 2, ...K$. And the Lagrangian is invariant
under the
 discrete symmetries. In addition, we require that above the GUT scale or 
including all the
KK states, the bulk should preserve the original GUT gauge symmetries
and supersymmetries, i. e., we can not project out all the KK states of
the fields in the theories, and the supersymmetric 
GUT models are broken down to the 4-dimensional $N=1$
 supersymmetric $SU(3)\times SU(2) \times U(1)^{n-3}$ model in the
bulk for the zero modes. 

Assume the local coordinates for extra dimensions in
the $I-th$ special brane neighborhood are $y_I^1$, $y_I^2$,
..., $y_I^n$, the action of any element 
$\gamma_i^I$ $\subset$ $\Gamma_I$ on $U_I$ can be expressed as
\begin{eqnarray}
\gamma_i^I: ~~~( y_I^1, y_I^2, ..., y_I^n) \subset U_I \longrightarrow
 (\gamma_i^I y_I^1, 
\gamma_i^I y_I^2, ..., \gamma_i^I y_I^n)
\subset U_I ~,~\,
\end{eqnarray}
where the $I-th$ special brane position is the only fixed point, 
line, or hypersurface for the
whole group $\Gamma_I$ as long as the neighborhood is small enough. 

The Lagrangian is invariant under the discrete symmetry in
the neighborhood $M^4\times U_I$ of the $I-th$ special brane, 
i. e., for any element $\gamma_i^I$ $\subset$ $\Gamma_I$
\begin{eqnarray}
{\cal L} (x^{\mu}, \gamma_i^I y_I^1, \gamma_i^I y_I^2, ..., \gamma_i^I
y_I^n)
={\cal L} (x^{\mu},  y_I^1, y_I^2, ..., y_I^n) ~,~\,
\end{eqnarray}
where $ (y_I^1, y_I^2, ..., y_I^n) \subset U_I$. So, for a 
generic bulk multiplet $\Phi$
which fills a representation of bulk gauge group $G$, we have
\begin{eqnarray}
\Phi (x^{\mu}, \gamma_i^I y_I^1, \gamma_i^I y_I^2, ..., \gamma_i^I y_I^n)
=
 \eta_{\Phi}^I
(R_{\gamma_i^I})^{l_\Phi} \Phi (x^{\mu},  y_I^1, y_I^2, ..., y_I^n) 
(R_{\gamma_i^I}^{-1})^{m_\Phi}
~,~\,
\end{eqnarray} 
where $\eta^I_{\Phi}$ is an element of the discrete
symmetry and can be determined from the Lagrangian 
(up to an element in $\Gamma_I$ for the matter fields), 
$\l_{\Phi}$ and $m_{\Phi}$ are the non-negative integers
determined by the representation of $\Phi$
under the gauge group $G$.
 Moreover, $R_{\gamma_i^I}$ is
an element in $G$, and $R_{\Gamma_I}$ is a discrete subgroup of $G$. We
will 
choose $R_{\gamma_i^I}$ as 
the matrix representation for $\gamma_i^I$ in the adjoint representation
of 
gauge group $G$. The consistent condition for $R_{\gamma_i^I}$ is
\begin{eqnarray}
R_{\gamma_i^I} R_{\gamma_j^I} = R_{\gamma_i^I \gamma_j^I} ~,~ 
\forall \gamma_i^I,~\gamma_j^I \subset \Gamma_I ~.~\,
\end{eqnarray} 
Mathematical speaking, the map 
$R:~ \Gamma_I \longrightarrow R_{\Gamma_I} \subset G $ is
a homomorphism. 
Because the special branes are fixed under the
 discrete symmetry transformations,
the gauge group on the $I-th$ special brane is 
the subgroup of $G$ which commutes with
$R_{\Gamma_I}$, and we denote the subgroup 
as $G/R_{\Gamma_I}$. For the zero modes, 
the bulk gauge group is broken down to
 the subgroup of $G$ which commutes with all $R_{\Gamma_I}$,
 i. e., $R_{\Gamma_1}$,
$R_{\Gamma_2}$, ..., $R_{\Gamma_K}$, and we denote
the subgroup as $G/\{R_{\Gamma_1}, R_{\Gamma_2}, ...,
 R_{\Gamma_K}\}$. In addition, 
if the theory is supersymmetric,
the special branes will preserve part of the bulk supersymmetry, and the
zero modes in the bulk also preserve part of the supersymmetry, in other
words, the
supersymmetry can be broken on the special branes for all the modes,
 and in the bulk for the zero modes.

In addition, we only have the KK states which satisfy the local
 and global discrete  symmetries in the theories because the KK modes, 
which do not satisfy the local
and global discrete symmetries, are projected out under our ansatz.
Therefore, we can construct the theories with only zero modes because all
the
KK modes are projected out, or the theories which have large extra
dimensions
and arbitrarily heavy KK states for there is no simple relation between
the
mass scales of extra dimensions and the masses of KK states. By the
way, we are only interested in the second kind of scenarios.

\section{Discrete Symmetry on the Space-Time $M^4\times M^1$}
We would like to generalize our previous models~\cite{LIBLS} to the
models on the space-time $M^4\times S^1$, 
and $M^4\times I^1$. 
 We obtain that the general models on 
$M^4\times I^1$ can be obtained from the general models on
$M^4\times S^1$ by moduloing the $(Z_2)^{k_0}$ symmetry
in which $k_0$ is the positive integer.
And the models on $M^4\times S^1/(Z_2\times Z_2')$
~\cite{Nbhn1} is an special case for $k_0$=2 and no 3-branes in the bulk.

We assume 
the corresponding coordinates for the space-time are $x^{\mu}$, 
($\mu = 0, 1, 2, 3$),
$y\equiv x^5$, the radius for the circle $S^1$ is $R$,
and the length for the interval $I^1$ is $\pi R$. 
We also assume that there are some special
3-branes along the fifth dimension,
and there is a $Z_2$ reflection symmetry in each 3-brane neighborhood. 

\subsection{Discrete Symmetry on $M^4\times S^1$}
Assuming we have $n+1$ parallel 
3-branes along the $S^1$, and
their fifth coordinates are: $y_0=0 < y_1 < y_2 < ... < y_n < 2 \pi R$.
We define the local fifth coordinate for the $i$-th brane as
$y_i'\equiv y-y_i$, and then, $y_0'\equiv y$. 
In addition, the equivalent class
for the reflection $Z_2$ symmetry in the $i-th$ brane beighborhood
is $y_i' \sim -y_i'$. And for that $Z_2$ symmetry,
we  define the corresponding $Z_2$ operator $P_i$ for $i$=0, 1,2, ..., n,
whose
eigenvalue is $\pm 1$, i. e., for a generic field or function,
we have
\begin{eqnarray}
P_i \phi(x^{\mu}, y_i') =\pm \phi(x^{\mu}, y_i')
~.~\,
\end{eqnarray} 
By the way, $P_i^2=1$, so, $\{1, P_i\}$ forms a $Z_2$ group.

Because if $y_i/(2 \pi R)$ is an irrational number, we will project
out all the KK states, which can not satisfy our requirement. So, we 
assume 
\begin{eqnarray}
y_i = {p_i\over q_i} 2 \pi R,~~~~~~~~~~{\rm for}~~ i=1, 2, ..., n  
~,~\,
\end{eqnarray}
where $p_i$ and $q_i$ are relative prime positive integers.

Assume $L$ is the least common multiple for all $p_i+q_i$, i. e.,
\begin{eqnarray}
L \equiv [p_1+q_1, p_2+q_2, ..., p_n+q_n]
~.~\,
\end{eqnarray} 
By Unique Prime Factorization theorem, we obtain that
\begin{eqnarray}
L = 2^{k_0} s_1^{k_1} s_2^{k_{2}} ... s_j^{k_{j}}
~,~\,
\end{eqnarray}  
where $2 < s_1 < s_2 <... < s_j$,  
$k_0$ is the non-negative integer, $s_i$ is the prime number
and $k_i$ is the positive integer for $i=1, 2, ..., j$.
Moreover, we define the effective physical radius $r$ as
\begin{eqnarray}
r = \left\{ \begin{array}{lll}
4R/L  & ~~~~{\rm for}~ k_0 \ge 2~, \\ 
2R/L  & ~~~~{\rm for}~ k_0 = 1~, \\ 
R/L   &~~~~{\rm for}~  k_0=0 ~. \end{array} \right.
\end{eqnarray}  
For a generic bulk field $\phi$, we obtain the KK modes expansions
\begin{eqnarray}
  \phi_{++} (x^\mu, y) &=& 
      \sum_{n=0}^{\infty} \frac{1}{\sqrt{2^{\delta_{n,0}} \pi R}} 
      \phi^{(2n)}_{++}(x^\mu) \cos{2ny \over r}~,~\,
\end{eqnarray}
\begin{eqnarray}
  \phi_{+-} (x^\mu, y) &=& 
      \sum_{n=0}^{\infty} \frac{1}{\sqrt{\pi R}} 
      \phi^{(2n+1)}_{+-}(x^\mu) \cos{(2n+1)y \over r}~,~\,
\end{eqnarray}
\begin{eqnarray}
  \phi_{-+} (x^\mu, y) &=& 
      \sum_{n=0}^{\infty} \frac{1}{\sqrt{\pi R}} \,
      \phi^{(2n+1)}_{-+}(x^\mu) \sin{(2n+1)y \over r}~,~\,
\end{eqnarray}
\begin{eqnarray}
  \phi_{--} (x^\mu, y) &=& 
      \sum_{n=0}^{\infty} \frac{1}{\sqrt{\pi R}} 
      \phi^{(2n+2)}_{--}(x^\mu) \sin{(2n+2)y \over r}~,~\,
\end{eqnarray}
where $n$ is a non-negative integer.
The 4-dimensional fields $\phi^{(2n)}_{++}$, $\phi^{(2n+1)}_{+-}$, 
$\phi^{(2n+1)}_{-+}$ and $\phi^{(2n+2)}_{--}$ acquire masses 
$2n/r$, $(2n+1)/r$, $(2n+1)/r$ and $(2n+2)/r$ upon the compactification.
Zero modes are contained only in $\phi_{++}$ fields,
thus, the matter content of massless sector is smaller
than that of the full 5-dimensional multiplet.
Moreover, because $0 < r \le R$,
 the masses of KK states ($n/r$)
can be set arbitrarily heavy if $L$ is large enough, i. e., 
we choose suitable $p_i$ and $q_i$ for some $i$, for
example, if $1/R$ is about TeV, $p_i=10^{13}-1$ and $q_i=10^{13}+1$,
we obtain that $1/r$ is at least about $10^{16}$ GeV, which is the usual
GUT scale.
Therefore, there is no simple relation between the physical size of 
the fifth dimension and the mass scales of KK modes.

For $k_0=0$ and $k_0=1$, we obtain that
\begin{eqnarray}
P_i \phi_{+\pm} (x^\mu, y) = \phi_{+\pm} (x^\mu, y)~,~\,
\end{eqnarray}
 \begin{eqnarray}
P_i \phi_{-\pm} (x^\mu, y) = -\phi_{-\pm} (x^\mu, y)~,~\,
\end{eqnarray}
for all $i=0, 1, 2, ..., n$. So, we only have one non-equivalent
$Z_2$ symmetry. And for $i=0$, we always have above equations
for $P_0$.

And for $k_0 \ge 2$, if $(p_i + q_i)$ is a multiple of $2^{k_0}$, i. e.,
 $2^{k_0} | (p_i+q_i)$, we obtain that
\begin{eqnarray}
P_i \phi_{\pm +} (x^\mu, y) = \phi_{\pm +} (x^\mu, y)~,~\,
\end{eqnarray}
 \begin{eqnarray}
P_i \phi_{\pm -} (x^\mu, y) = -\phi_{\pm -} (x^\mu, y)~,~\,
\end{eqnarray}
and if  $(p_i + q_i)$ is not a multiple of $2^{k_0}$, 
i. e., $2^{k_0} \not\vert (p_i+q_i)$, we obtain that
\begin{eqnarray}
P_i \phi_{+\pm} (x^\mu, y) = \phi_{+\pm} (x^\mu, y)~,~\,
\end{eqnarray}
 \begin{eqnarray}
P_i \phi_{-\pm} (x^\mu, y) = -\phi_{-\pm} (x^\mu, y)~.~\,
\end{eqnarray}
 So, we  have two non-equivalent $Z_2$ symmetries.

Because we need discrete symmetry to break the bulk gauge symmetry
and supersymmetry, we will concentrate on the scenario with
$k_0 \ge 2$. To be explicit, we would like to give two examples:
(I) $n=1$ and $4|(p_1+q_1)$, in this simple case,
 we can have two local $Z_2$ symmetries; (II) Suppose 
$n=3$, $4|(p_1+q_1)$, $p_2=q_2=1$,
$p_3=q_1$, and $q_3=p_1$, we have one global $Z_2$ symmetry and
one local $Z_2$ symmetry. And the local $Z_2$ symmetry will become
global if $p_1=1$ and $q_1=3$. This is the scenario discussed
in Ref.~\cite{LIBLS} where the global $Z_2$ symmetry has been moduloed
from the
manifold.

Furthermore, if we require that the models 
have one global $Z_2$ symmetry, then,
modulo this global $Z_2$ symmetry, we obtain the models with discrete
symmetry on the space-time $M^4\times S^1/Z_2$. 
And the two $Z_2$ symmetries in the two
 boundary 3-branes' neighborhoods are equivalent.
Let us explain this
in detail: suppose we have $2n+2$ special 3-branes, we require that
$y_0=0$,
$y_{n+1} = \pi R$, $p_i=q_{2n+2-i}$ and $q_i=p_{2n+2-i}$ where $i=$ 1, 2,
..., $n$, we will have one global $Z_2$ symmetry in which the equivalent
class
is $y \sim -y$. Moduloing this equivalent class, we obtain the models
on  $M^4\times S^1/Z_2$.

In general, if we require that the models have global $(Z_2)^{k_0}$
symmetry
for $k_0 > 1$,
 then, modulo the global $(Z_2)^{k_0}$ symmetry, we obtain the
models with discrete symmetry on the space-times $M^4\times
S^1/(Z_2)^{k_0}$.
And the two $Z_2$ symmetries in the two
 boundary 3-branes' neighborhoods are not equivalent.
As an example, we discuss the models with $k_0=2$.
Suppose we have $4n+4$ 3-branes, $y_{n+1}=\pi R/2$, $y_{2n+2}=\pi R$,
$y_{3n+3}= 3\pi R/2$. And for $i=1, 2, ..., n$, we have
$p_i=q_{4n+4-i}$, $q_i=p_{4n+4-i}$, $p_{2n+2-i}=p_i'$,
$q_{2n+2-i}=q_i'$, $p_{2n+2+i}=q_i'$,
$q_{2n+2+i}=p_i'$, where $p_i'$ and $q_i'$ are relative prime positive
integer and satisfy the equation
\begin{eqnarray}
{{p_i'}\over {q_i'}} = {{q_i-p_i}\over\displaystyle {2(p_i+q_i)}}
~.~\,
\end{eqnarray} 
If $n=0$, we obtain the models on $M^4\times S^1/(Z_2\times
Z_2')$~\cite{Nbhn1}.

\subsection{Discrete Symmetry on $M^4\times I^1$}
In this subsection, we would like to consider the discrete symmetry
on the space-time $M^4\times I^1$. Assume on two boundary 3-branes,
the fields should satisfy the Dirichlet or Neumann boundary condition,
we show that the general models on the space-time $M^4\times I^1$,
 contain the models on $M^4\times S^1/Z_2$ and
the models on $M^4\times S^1/(Z_2)^{k_0}$ for $k_0 > 1$.
Assuming we have $n+2$ parallel 3-branes along the $I^1$, and
their fifth coordinates are: $y_0=0 < y_1 < y_2 < ... < y_{n+1} = \pi R$.
We define the local fifth coordinate for the $i$-th brane as
$y_i'\equiv y-y_i$. And the equivalent class
for the reflection $Z_2$ symmetry in the $i-th$ brane beighborhood
is $y_i' \sim -y_i'$. Moreover, for that $Z_2$ symmetry,
we  define the corresponding $Z_2$ operator $P_i$ for 
$i$=0, 1,2, ..., n,  whose
eigenvalue is $\pm 1$, i. e., for a generic field or function,
we have
\begin{eqnarray}
P_i \phi(x^{\mu}, y_i') =\pm \phi(x^{\mu}, y_i')
~.~\,
\end{eqnarray} 

Because if $y_i/(\pi R)$ is an irrational number, we will project
out all the KK states, which can not satisfy our requirement. So, we 
assume 
\begin{eqnarray}
y_i = {p_i\over q_i} \pi R,~~~~~~~~~~ {\rm for}~~ i=1, 2, ..., n  
~,~\,
\end{eqnarray}
where $p_i$ and $q_i$ are relative prime positive integers.

Assume $L$ is the least common multiple for all $p_i+q_i$, i. e.,
\begin{eqnarray}
L \equiv [p_1+q_1, p_2+q_2, ..., p_n+q_n]
~.~\,
\end{eqnarray} 
By Unique Prime Factorization theorem, we obtain that
\begin{eqnarray}
L = 2^{l_0} s_1^{l_1} s_2^{l_{2}} ... s_j^{l_{j}}
~,~\,
\end{eqnarray}  
where $2 < s_1 < s_2 <... < s_j$,  
$l_0$ is the non-negative integer, $s_i$ is the prime number
and $l_i$ is the positive integer for $i=1, 2, ..., j$.
For the models on the space-time $M^4\times S^1/Z_2$,
we define the effective physical radius $r$ as
\begin{eqnarray}
r = \left\{ \begin{array}{ll}
2R/L  & ~~~~{\rm for}~ l_0 \ge 1~, \\ 
R/L   &~~~~{\rm for}~  l_0=0 ~. \end{array} \right.
\end{eqnarray}  

For $l_0=0$, we can still define the effective radius as
\begin{eqnarray}
r ={{2R}\over L} ~.~\,
\end{eqnarray} 
This kind of the models can not be obtained from the models in
the last subsection by moduloing one global $Z_2$ symmetry, however,
they can be obtained from the models in the last subesection
by moduloing global $(Z_2)^{k_0}$ symmetries for $k_0 > 1$. 

For a generic bulk field $\phi$, we obtain the KK modes expansions
\begin{eqnarray}
  \phi_{++} (x^\mu, y) &=& 
      \sum_{n=0}^{\infty} \frac{1}{\sqrt{2^{\delta_{n,0}} \pi R}} 
      \phi^{(2n)}_{++}(x^\mu) \cos{2ny \over r}~,~\,
\end{eqnarray}
\begin{eqnarray}
  \phi_{+-} (x^\mu, y) &=& 
      \sum_{n=0}^{\infty} \frac{1}{\sqrt{\pi R}} 
      \phi^{(2n+1)}_{+-}(x^\mu) \cos{(2n+1)y \over r}~,~\,
\end{eqnarray}
\begin{eqnarray}
  \phi_{-+} (x^\mu, y) &=& 
      \sum_{n=0}^{\infty} \frac{1}{\sqrt{\pi R}} \,
      \phi^{(2n+1)}_{-+}(x^\mu) \sin{(2n+1)y \over r}~,~\,
\end{eqnarray}
\begin{eqnarray}
  \phi_{--} (x^\mu, y) &=& 
      \sum_{n=0}^{\infty} \frac{1}{\sqrt{\pi R}} 
      \phi^{(2n+2)}_{--}(x^\mu) \sin{(2n+2)y \over r}~,~\,
\end{eqnarray}
where $n$ is a non-negative integer.
The 4-dimensional fields $\phi^{(2n)}_{++}$, $\phi^{(2n+1)}_{+-}$, 
$\phi^{(2n+1)}_{-+}$ and $\phi^{(2n+2)}_{--}$ acquire masses 
$2n/r$, $(2n+1)/r$, $(2n+1)/r$ and $(2n+2)/r$ upon the compactification.
And the zero modes are contained only in $\phi_{++}$ fields.
Moreover, because $0 < r \le R$,
 the masses of KK states ($n/r$)
can be set arbitrarily heavy if $L$ is large enough,
i. e.,  we choose suitable $p_i$ and $q_i$, for some $i$.
So, there is no simple relation between the physical size of 
the fifth dimension and the mass scales of KK states.

(I) First, we discuss the models on the space-times $M^4\times S^1/Z_2$.
We should keep in mind that there is one global $Z_2$ symmetry has been
moduloed
from $S^1$, which we can call it as $P_0$ or $P_{n+1}$.

For $l_0=0$ , we obtain that
\begin{eqnarray}
P_i \phi_{+\pm} (x^\mu, y) = \phi_{+\pm} (x^\mu, y)~,~\,
\end{eqnarray}
 \begin{eqnarray}
P_i \phi_{-\pm} (x^\mu, y) = -\phi_{-\pm} (x^\mu, y)~,~\,
\end{eqnarray}
for all $i= 1, 2, ..., n$. Under our assumption, those $Z_2$ symmetry
are equivalent to the global $Z_2$ symmetry, so, we just have one
independet $Z_2$ symmetry.

And for $l_0 \ge 1$, if $(p_i + q_i)$ is a multiple of $2^{l_0}$, i. e.,
 $2^{l_0} | (p_i+q_i)$, we obtain that
\begin{eqnarray}
P_i \phi_{\pm +} (x^\mu, y) = \phi_{\pm +} (x^\mu, y)~,~\,
\end{eqnarray}
 \begin{eqnarray}
P_i \phi_{\pm -} (x^\mu, y) = -\phi_{\pm -} (x^\mu, y)~,~\,
\end{eqnarray}
this $Z_2$ symmetry is not equivalent to the global symmetry.
And if  $(p_i + q_i)$ is not a multiple of $2^{l_0}$,
i. e., $2^{l_0} \not\vert (p_i+q_i)$,
 we obtain that
\begin{eqnarray}
P_i \phi_{+\pm} (x^\mu, y) = \phi_{+\pm} (x^\mu, y)~,~\,
\end{eqnarray}
 \begin{eqnarray}
P_i \phi_{-\pm} (x^\mu, y) = -\phi_{-\pm} (x^\mu, y)~,~\,
\end{eqnarray}
this $Z_2$ symmetry is equivalent to the global $Z_2$ symmetry.
In short,
 we  have two independent $Z_2$ symmetries, in which one can be considered
as global $Z_2$ symmetry.

Because we need discrete symmetry to break the bulk gauge symmetry
and supersymmetry, we will concentrate on the scenario with
$l_0 \ge 1$. To be explicit, we would like to give one examples:
 $n=1$ and $2|(p_1+q_1)$. In this simple case,
 we can have one local $Z_2$ symmetries and one global $Z_2$ symmetry. 
And that local $Z_2$ symmetry becomes global if 
$p_1=q_1=1$.

(II) If $l_0=0$ and $r=2R/L$,
this kind of models can  be obtained from the models in
the last subsection by moduloing the global $(Z_2)^{k_0}$ symmetries
for $k_0~>~1$, so,
$P_0$ and $P_{n+1}$ are two non-equivalent
$Z_2$ symmetries. If at start point, the original extra space manifold is
$I^1$, we can consider that  
there are no global $Z_2$ symmetries for $P_0$ and $P_{n+1}$.

Because $l_0=0$, there are two cases: $p_i$ is odd and $q_i$ is even,
or $p_i$ is even and $q_i$ is odd. If $p_i$ is even,
\begin{eqnarray}
P_i \phi_{+\pm} (x^\mu, y) = \phi_{+\pm} (x^\mu, y)~,~\,
\end{eqnarray}
 \begin{eqnarray}
P_i \phi_{-\pm} (x^\mu, y) = -\phi_{-\pm} (x^\mu, y)~,~\,
\end{eqnarray}
and if $p_i$ is odd
\begin{eqnarray}
P_i \phi_{\pm +} (x^\mu, y) = \phi_{\pm +} (x^\mu, y)~,~\,
\end{eqnarray}
 \begin{eqnarray}
P_i \phi_{\pm -} (x^\mu, y) = -\phi_{\pm -} (x^\mu, y)~.~\,
\end{eqnarray}
Therefore, we can have two local $Z_2$ symmetries, which can be
thought as global symmetries if we consider the original
manifold for the extra dimension is $S^1$.

\section{GUT Breaking on the Space-Time $M^4\times M^1$}
In this section,
 we would like to discuss the supersymmetric $SU(5)$ model on the
space-time 
$M^4\times M^1$ with two discrete $Z_2$ symmetries. 
We assume that the $SU(5)$ gauge fields and two 5-plet Higgs
hypermultiplets
in the bulk, and the Standard Model fermions can be on the 3-brane or in
the
bulk. 

As we know, the $N=1$ supersymmetric theory in 5-dimension have 
8 real supercharges,
corresponding to $N=2$ supersymmetry in 4-dimension. The vector multiplet
physically contains a vector boson $A_M$ where $M=0, 1, 2, 3, 5$, 
two Weyl gauginos $\lambda_{1,2}$, and a real scalar $\sigma$. 
In terms of 4-dimensional
$N=1$ language, it contains a vector multiplet $V(A_{\mu}, \lambda_1)$ and
a chiral multiplet $\Sigma((\sigma+iA_5)/\sqrt 2, \lambda_2)$ which
transform
in the adjoint representation of $SU(5)$.
And the 5-dimensional hypermultiplet physically has two complex scalars
$\phi$ and $\phi^c$, a Dirac fermion $\Psi$, and can be decomposed into 
two 4-dimensional chiral mupltiplets $\Phi(\phi, \psi \equiv \Psi_R)$
and $\Phi^c(\phi^c, \psi^c \equiv \Psi_L)$, which transform as
conjugate representations
of each other under the gauge group. For instance,
 we have two Higgs chiral multiplets $H_u$ and $H_d$, which transform 
as $ 5$ and $\bar 5$ under SU(5) gauge symmetry, and their
mirror $H_u^c$ and $H_d^c$, which transform as $ \bar 5$ and $ 
5$ under SU(5) gauge symmetry.

The general action for the $SU(5)$ gauge fields and their couplings to the
bulk hypermultiplet $\Phi$ is~\cite{NAHGW} 
\begin{eqnarray}
S&=&\int{d^5x}\frac{1}{k g^2}
{\rm Tr}\left[\frac{1}{4}\int{d^2\theta} \left(W^\alpha W_\alpha+{\rm H.
C.}\right)
\right.\nonumber\\&&\left.
+\int{d^4\theta}\left((\sqrt{2}\partial_5+ {\bar \Sigma })
e^{-V}(-\sqrt{2}\partial_5+\Sigma )e^V+
\partial_5 e^{-V}\partial_5 e^V\right)\right]
\nonumber\\&&
+\int{d^5x} \left[ \int{d^4\theta} \left( {\Phi}^c e^V {\bar \Phi}^c +
{\bar \Phi} e^{-V} \Phi \right)
\right.\nonumber\\&&\left.
+ \int{d^2\theta} \left( {\Phi}^c (\partial_5 -{1\over {\sqrt 2}} \Sigma)
\Phi + {\rm H. C.}
\right)\right]~.~\,
\end{eqnarray}

Because the action is invariant under the parity
 $P_i$,
 we obtain that under the parity operator $P_i$, the
vector multiplet transforms as
\begin{eqnarray}
V(x^{\mu},y_i')&\to  V(x^{\mu},-y_i') = P_i V(x^{\mu}, y_i') P_i^{-1}
~,~\,
\end{eqnarray}
\begin{eqnarray}
 \Sigma(x^{\mu},y_i') &\to\Sigma(x^{\mu},-y_i') = - P_i \Sigma(x^{\mu},
y_i') P_i^{-1}
~,~\,
\end{eqnarray}
if the hypermultiplet $\Phi$ is a $5$ or $\bar 5$ $SU(5)$ multiplet, we
have 
\begin{eqnarray}
\Phi(x^{\mu},y_i')&\to \Phi(x^{\mu}, -y_i')  = \eta_{\Phi} P_i
\Phi(x^{\mu},y_i')
~,~\,
\end{eqnarray}
\begin{eqnarray}
\Phi^c(x^{\mu},y_i') &\to \Phi^c(x^{\mu}, -y_i')  = -\eta_{\Phi} P_i
\Phi^c(x^{\mu},y_i')
~,~\,
\end{eqnarray}
and if the hypermultiplet $\Phi$ is a $10$ or $\bar {10}$ $SU(5)$
multiplet, we have
\begin{eqnarray}
\Phi(x^{\mu},y_i')&\to \Phi(x^{\mu}, -y_i')  = \eta_{\Phi} P_i
\Phi(x^{\mu},y_i') P_i^{-1}
~,~\,
\end{eqnarray}
\begin{eqnarray}
\Phi^c(x^{\mu},y_i') &\to \Phi^c(x^{\mu}, -y_i')  = -\eta_{\Phi} P_i
\Phi^c(x^{\mu},y_i') P_i^{-1}
~,~\,
\end{eqnarray}
where $\eta_{\Phi} = \pm 1$. 

For simplicity, let us denote the two non-equivalent $Z_2$ symmetries as
$P$ and $P'$.
We choose the following matrix representations for the parities $P$ and 
 $P'$ which are expressed in the adjoint representaion of 
$SU(5)$
\begin{equation}
P={\rm diag}(+1, +1, +1, +1, +1)~,~P'={\rm diag}(-1, -1, -1, +1, +1)
 ~.~\,
\end{equation}
So, upon the parity $P'$,
 the gauge generators $T^A$ where A=1, 2, ..., 24 for $SU(5)$
are separated into two sets: $T^a$ are the gauge generators for
the Standard Model gauge group, and $T^{\hat a}$ are the other broken
gauge generators 
\begin{equation}
P~T^a~P^{-1}= T^a ~,~ P~T^{\hat a}~P^{-1}= T^{\hat a}
~,~\,
\end{equation}
\begin{equation}
P'~T^a~P^{'-1}= T^a ~,~ P'~T^{\hat a}~P^{'-1}= - T^{\hat a}
~.~\,
\end{equation}

Choosing $\eta_{H_u}=+1$ and $\eta_{H_d}=+1$, we obtain the 
 particle spectra, which are given in 
Table 1. The bulk 4-dimensional $N=2$ supersymmetry and 
$SU(5)$ gauge symmetry are broken down to the 4-dimensional $N=1$ 
supersymmetry and
$SU(3)\times SU(2)\times U(1)$ gauge symmetry in the bulk for the zero
modes,
and on the special 3-branes
which preserve $Z_2$ symmetry $P'$ for all the modes.
Including the KK states, the gauge symmetry
 on the special 3-branes, which preserve $Z_2$ symmetry $P$,
is $SU(5)$. In addition,
the 4-dimensional supersymmetry on the 3-branes
 is ${1/2}$ of the bulk 4-dimensional supersymmetry or $N=1$ due
to the $Z_2$ symmetry in the brane neighbohood.
 Moreover, the Standard Model fermions can
be in the bulk or on the 3-brane, and the discussion are similar to 
those in Ref.~[4-8], so, we will not repeat them here.

By the way, 
one can also discuss the non-supersymmetric $SU(6)$ and $SO(10)$ breaking,
however, there are zero modes for $A_5^{\hat a}$ where ${\hat a}$ is the
index
related to the broken gauge generators under two $Z_2$ symmetries.

\renewcommand{\arraystretch}{1.4}
\begin{table}[t]
\caption{Parity assignment and masses ($n\ge 0$) of the fields in the
SU(5) 
 gauge and Higgs multiplets.
The indices $F$, $T$ are for doublet and triplet, respectively. 
\label{tab:chiral}}
\vspace{0.4cm}
\begin{center}
\begin{tabular}{|c|c|c|}
\hline        
$(P,P')$ & field & mass\\ 
\hline
$(+,+)$ &  $V^a_{\mu}$, $H^F_u$, $H^F_d$ & ${{2n}\over r}$ \\
\hline
$(+,-)$ &  $V^{\hat{a}}_{\mu}$,  $H^T_u$, $H^T_d$ & ${{2n+1}\over r}$  \\
\hline
$(-,+)$ &  $\Sigma^{\hat{a}}$, ${H}^{cT}_u$, ${H}^{cT}_d$  & ${{2n+1}\over
r}$ \\
\hline
$(-,-)$ &  $\Sigma^a$, ${ H}^{cF}_u$, ${H}^{cF}_d$ &  ${{2n+2}\over r}$\\
\hline
\end{tabular}
\end{center}
\end{table}

\section{Discrete Symmetry on the Space-Time $M^4\times M^1\times M^1$}
We would like to discuss the
models where there are some parallel 4-branes with $Z_2$ reflection
symmetry
along the fifth and sixth dimensions
 on the space-time $M^4\times M^1 \times M^1$, 
in which $M^1$ can be $S^1$, $S^1/Z_2$, and $I^1$.
Because the
extra space manifold is the product of two 1-dimensional manifold,
 we only discuss the models
on the space-times $M^4\times S^1\times S^1$
 as a representative because the
discussions of the models with other combinations are similar.
In addition, we discuss the models on the space-time
$M^4\times S^1/Z_2 \times S^1/Z_2$ where there are some 3-branes
with $Z_2$ symmetry in the bulk.

The corresponding coordinates for the space-time are $x^{\mu}$, 
($\mu = 0, 1, 2, 3$), $y\equiv x^5$, $z\equiv x^6$, and
the radii for the $y$ and $z$ directions are $R_1$ and $R_2$,
respectively.

\subsection{Discrete Symmetry on $M^4\times S^1 \times S^1$}
First, we would like to discuss the discrete symmetry on the space-time 
$M^4\times S^1 \times S^1$. Assume that along the $y$ and $z$ directions,
we
have $n+1$ and $m+1$ parallel 4-branes with $Z_2$ reflection symmetry,
respectively. The 5-th coordinates for the parallel
4-branes along the $y$ direction are $y_0=0 < y_1 < y_2 < ... < y_n < 2
\pi R_1$,
and the 6-th coordinates for the parallel
4-branes along the $z$ direction are $z_0=0 < z_1 < z_2 < ... < z_m < 2
\pi R_2$.

We denote the local coordinate for the $i$-th 4-brane along the 
$y$ direction  as
$y_i'\equiv y-y_i$. In addition, for the $Z_2$ symmetry in 
the $i-th$ 4-brane neighborhood, 
we define a $Z_2$ operator $P_i^y$ for $i$=0, 1,2, ..., n,  whose
eigenvalue is $\pm 1$, i. e., for a generic field or function,
we have
\begin{eqnarray}
P_i^y \phi(x^{\mu}, y_i', z) =\pm \phi(x^{\mu}, y_i', z)
~.~\,
\end{eqnarray} 

Similarly, we denote the local coordinate for the $i$-th 4-brane along the 
$z$ direction  as
$z_i'\equiv z-z_i$. And for the $Z_2$ symmetry in the $i-th$ 4-brane
neighborhood,
we define a $Z_2$ operator $P_i^z$ for $i$=0, 1,2, ..., m 
\begin{eqnarray}
P_i^z \phi(x^{\mu}, y, z_i') =\pm \phi(x^{\mu}, y, z_i')
~.~\,
\end{eqnarray} 

Because if $y_i/(2 \pi R_1)$ or $z_i/(2 \pi R_1)$
 is an irrational number, we will project
out all the KK states, which can not satisfy our requirement. So, we 
assume 
\begin{eqnarray}
y_i = {p^y_i\over q^y_i} 2 \pi R_1, ~~~~~~~~~~ {\rm for}~~ i=1, 2, ..., n  
~,~\,
\end{eqnarray}
\begin{eqnarray}
z_i = {p^z_i\over q^z_i} 2 \pi R_2, ~~~~~~~~~~ {\rm for}~~ i=1, 2, ..., m 
~,~\,
\end{eqnarray}
where $p^y_i$ and $q^y_i$ are relative prime positive integers, and
$p^z_i$ and $q^z_i$ are relative prime positive integers.

Assume $L_y$ is the least common multiple for all $p_i^y+q_i^y$, 
and $L_z$ is  the least common multiple for all $p_i^z+q_i^z$, i. e.,
\begin{eqnarray}
L_y \equiv [p_1^y+q_1^y, p_2^y+q_2^y, ..., p_n^y+q_n^y]
~,~\,
\end{eqnarray} 
\begin{eqnarray}
L_z \equiv [p_1^z+q_1^z, p_2^z+q_2^z, ..., p_m^z+q_m^z]
~.~\,
\end{eqnarray}

By Unique Prime Factorization theorem, we obtain that
\begin{eqnarray}
L_y = 2^{k_0} s_1^{k_1} s_2^{k_{2}} ... s_u^{k_{u}}
~,~\,
\end{eqnarray} 
\begin{eqnarray}
L_z = 2^{l_0} t_1^{l_1} t_2^{l_{2}} ... t_v^{l_{v}}
~,~\,
\end{eqnarray} 
 
where $2 < s_1 < s_2 <... < s_u$,  $2 < t_1 < t_2 <... < t_v$,
$k_0$ and $l_0$
are the non-negative integers, $s_i$ and $t_j$ are the prime numbers,
and $k_i$ and $l_j$ are the positive integers for $i=1, 2, ..., u$
and $j=1, 2, ..., v$.
And we define the effective physical radii $r_1$ and $r_2$ as
\begin{eqnarray}
r_1 = \left\{ \begin{array}{lll}
4R_1/L_y  & ~~~~{\rm for}~ k_0 \ge 2~, \\ 
2R_1/L_y  & ~~~~{\rm for}~ k_0 = 1~, \\ 
R_1/L_y   &~~~~{\rm for}~  k_0=0 ~, \end{array} \right.
\end{eqnarray}  
\begin{eqnarray}
r_2 = \left\{ \begin{array}{lll}
4R_1/L_z  & ~~~~{\rm for}~ l_0 \ge 2~, \\ 
2R_1/L_z  & ~~~~{\rm for}~ l_0 = 1~, \\ 
R_1/L_z   &~~~~{\rm for}~  l_0=0 ~. \end{array} \right.
\end{eqnarray}  

For a generiac bulk field $\phi$, we obtain the KK modes expansions
\begin{eqnarray}
\phi_{++++}(x^{\mu}, y, z) = \sum_{n=0}^{\infty} \sum_{m=0}^{\infty}
\phi_{++++}^{(2n, 2m)}(x^{\mu}) A_{++}^{2n} (y, r_1) A_{++}^{2m} (z, r_2) 
 ~,~\,
\end{eqnarray}
\begin{eqnarray}
\phi_{+++-}(x^{\mu}, y, z) = \sum_{n=0}^{\infty} \sum_{m=0}^{\infty}
\phi_{+++-}^{(2n, 2m+1)}(x^{\mu}) A_{++}^{2n} (y, r_1) A_{+-}^{2m+1} (z,
r_2) 
 ~,~\,
\end{eqnarray}
\begin{eqnarray}
\phi_{++-+}(x^{\mu}, y, z) = \sum_{n=0}^{\infty} \sum_{m=0}^{\infty}
\phi_{++-+}^{(2n, 2m+1)}(x^{\mu}) A_{++}^{2n} (y, r_1) A_{-+}^{2m+1} (z,
r_2) 
 ~,~\,
\end{eqnarray}
\begin{eqnarray}
\phi_{++--}(x^{\mu}, y, z) = \sum_{n=0}^{\infty} \sum_{m=0}^{\infty}
\phi_{++--}^{(2n, 2m+2)}(x^{\mu}) A_{++}^{2n} (y, r_1) A_{--}^{2m+2} (z,
r_2) 
 ~,~\,
\end{eqnarray}
\begin{eqnarray}
\phi_{+-++}(x^{\mu}, y, z) = \sum_{n=0}^{\infty} \sum_{m=0}^{\infty}
\phi_{+-++}^{(2n+1, 2m)}(x^{\mu}) A_{+-}^{2n+1} (y, r_1) A_{++}^{2m} (z,
r_2) 
 ~,~\,
\end{eqnarray}
\begin{eqnarray}
\phi_{+-+-}(x^{\mu}, y, z) = \sum_{n=0}^{\infty} \sum_{m=0}^{\infty}
\phi_{+-+-}^{(2n+1, 2m+1)}(x^{\mu}) A_{+-}^{2n+1} (y, r_1) A_{+-}^{2m+1}
(z, r_2) 
 ~,~\,
\end{eqnarray}
\begin{eqnarray}
\phi_{+--+}(x^{\mu}, y, z) = \sum_{n=0}^{\infty} \sum_{m=0}^{\infty}
\phi_{+--+}^{(2n+1, 2m+1)}(x^{\mu}) A_{+-}^{2n+1} (y, r_1) A_{-+}^{2m+1}
(z, r_2) 
 ~,~\,
\end{eqnarray}
\begin{eqnarray}
\phi_{+---}(x^{\mu}, y, z) = \sum_{n=0}^{\infty} \sum_{m=0}^{\infty}
\phi_{+---}^{(2n+1, 2m+2)}(x^{\mu}) A_{+-}^{2n+1} (y, r_1) A_{--}^{2m+2}
(z, r_2) 
 ~,~\,
\end{eqnarray}
\begin{eqnarray}
\phi_{-+++}(x^{\mu}, y, z) = \sum_{n=0}^{\infty} \sum_{m=0}^{\infty}
\phi_{-+++}^{(2n+1, 2m)}(x^{\mu}) A_{-+}^{2n+1} (y, r_1) A_{++}^{2m} (z,
r_2) 
 ~,~\,
\end{eqnarray}
\begin{eqnarray}
\phi_{-++-}(x^{\mu}, y, z) = \sum_{n=0}^{\infty} \sum_{m=0}^{\infty}
\phi_{-++-}^{(2n+1, 2m+1)}(x^{\mu}) A_{-+}^{2n+1} (y, r_1) A_{+-}^{2m+1}
(z, r_2) 
 ~,~\,
\end{eqnarray}
\begin{eqnarray}
\phi_{-+-+}(x^{\mu}, y, z) = \sum_{n=0}^{\infty} \sum_{m=0}^{\infty}
\phi_{-+-+}^{(2n+1, 2m+1)}(x^{\mu}) A_{-+}^{2n+1} (y, r_1) A_{-+}^{2m+1}
(z, r_2) 
 ~,~\,
\end{eqnarray}
\begin{eqnarray}
\phi_{-+--}(x^{\mu}, y, z) = \sum_{n=0}^{\infty} \sum_{m=0}^{\infty}
\phi_{-+--}^{(2n+1, 2m+2)}(x^{\mu}) A_{-+}^{2n+1} (y, r_1) A_{--}^{2m+2}
(z, r_2) 
 ~,~\,
\end{eqnarray}
\begin{eqnarray}
\phi_{--++}(x^{\mu}, y, z) = \sum_{n=0}^{\infty} \sum_{m=0}^{\infty}
\phi_{--++}^{(2n+2, 2m)}(x^{\mu}) A_{--}^{2n+2} (y, r_1) A_{++}^{2m} (z,
r_2) 
 ~,~\,
\end{eqnarray}
\begin{eqnarray}
\phi_{--+-}(x^{\mu}, y, z) = \sum_{n=0}^{\infty} \sum_{m=0}^{\infty}
\phi_{--+-}^{(2n+2, 2m+1)}(x^{\mu}) A_{--}^{2n+2} (y, r_1) A_{+-}^{2m+1}
(z, r_2) 
 ~,~\,
\end{eqnarray}
\begin{eqnarray}
\phi_{---+}(x^{\mu}, y, z) = \sum_{n=0}^{\infty} \sum_{m=0}^{\infty}
\phi_{---+}^{(2n+2, 2m+1)}(x^{\mu}) A_{--}^{2n+2} (y, r_1) A_{-+}^{2m+1}
(z, r_2) 
 ~,~\,
\end{eqnarray}
\begin{eqnarray}
\phi_{----}(x^{\mu}, y, z) = \sum_{n=0}^{\infty} \sum_{m=0}^{\infty}
\phi_{----}^{(2n+2, 2m+2)}(x^{\mu}) A_{--}^{2n+2} (y, r_1) A_{--}^{2m+2}
(z, r_2) 
 ~,~\,
\end{eqnarray}
where
\begin{eqnarray}
A_{++}^{2n}(y, r_1) = {1 \over\displaystyle {\sqrt{2^{\delta_{n, 0}} \pi
R_1 }}}
 \cos{{2n y}\over {r_1}}
~,~\,
\end{eqnarray}
\begin{eqnarray}
A_{+-}^{2n+1}(y, r_1) = {1 \over\displaystyle {\sqrt{ \pi R_1 }}}
 \cos{{(2n+1)y}\over {r_1}}
~,~\,
\end{eqnarray}
\begin{eqnarray}
A_{-+}^{2n+1}(y, r_1) = {1 \over\displaystyle {\sqrt{ \pi R_1 }}}
 \sin{{(2n+1)y}\over {r_1}}
~,~\,
\end{eqnarray}
\begin{eqnarray}
A_{--}^{2n+2}(y, r_1) = {1 \over\displaystyle {\sqrt{ \pi R_1 }}}
 \sin{{(2n+2)y}\over {r_1}}
~.~\,
\end{eqnarray}
Similarly, we define $A_{++}^{2n}(z, r_2)$, $A_{+-}^{2n+1}(z, r_2)$,
$A_{-+}^{2n+1}(z, r_2)$, $A_{--}^{2n+2}(z, r_2)$.

The 4-dimensional fields $\phi^{(n, m)}$ acquire masses 
$\sqrt {n^2/r_1^2+ m^2/r_2^2}$ upon the compactification.
And the
zero modes are contained only in $\phi_{++++}$ fields.
Moreover, because $0 < r_1 \le R_1$ and $0 < r_2 \le R_2$,
 the masses of KK states ($\sqrt {n^2/r_1^2+ m^2/r_2^2}$)
can be set arbitrarily heavy if $L_y$ and $L_z$ are
large enough, i. e.,  we choose suitable $(p_i^y, q_i^y)$ for some $i$,
and $(p_j^z, q_j^z)$ for some $j$.
So, there is no simple relation between the physical size of 
the extra dimensions and the mass scales of KK modes.

For $k_0=0$ and $k_0=1$, we obtain that
\begin{eqnarray}
P_i^y \phi_{+\pm \pm \pm} (x^\mu, y, z) = \phi_{+\pm \pm \pm} (x^\mu, y,
z)~,~\,
\end{eqnarray}
 \begin{eqnarray}
P_i^y \phi_{-\pm \pm \pm } (x^\mu, y, z) = -\phi_{-\pm \pm \pm} (x^\mu, y,
z)~,~\,
\end{eqnarray}
for all $i=0, 1, 2, ..., n$. So, we only have one independent
$Z_2$ symmetry. And for $i=0$, we always have above equations
for $P_0^y$.

Moreover, for $k_0 \ge 2$, if $(p_i^y + q_i^y)$ is a multiple of
$2^{k_0}$, i. e.,
 $2^{k_0} | (p_i^y+q_i^y)$, we obtain that
\begin{eqnarray}
P_i^y \phi_{\pm + \pm \pm} (x^\mu, y, z) = \phi_{\pm + \pm \pm} (x^\mu, y,
z)~,~\,
\end{eqnarray}
 \begin{eqnarray}
P_i^y \phi_{\pm - \pm \pm} (x^\mu, y, z) = -\phi_{\pm - \pm \pm} (x^\mu,
y, z)~,~\,
\end{eqnarray}
and if  $(p_i^y + q_i^y)$ is not a multiple of $2^{k_0}$,
i. e., $2^{k_0} \not\vert (p_i^y+q_i^y)$,
 we obtain that
\begin{eqnarray}
P_i^y \phi_{+\pm \pm \pm} (x^\mu, y, z) = \phi_{+\pm \pm \pm} (x^\mu, y,
z)~,~\,
\end{eqnarray}
 \begin{eqnarray}
P_i^y \phi_{-\pm \pm \pm} (x^\mu, y, z) = -\phi_{-\pm \pm \pm} (x^\mu, y,
z)~.~\,
\end{eqnarray}
 So, we  have two non-equivalent $Z_2$ symmetries along the $y$ direction.

Similarly, for $l_0=0$ and $l_0=1$, we obtain that
\begin{eqnarray}
P_i^z \phi_{\pm \pm + \pm} (x^\mu, y, z) = \phi_{\pm \pm + \pm} (x^\mu, y,
z)~,~\,
\end{eqnarray}
 \begin{eqnarray}
P_i^z \phi_{\pm \pm - \pm } (x^\mu, y, z) = -\phi_{\pm \pm - \pm} (x^\mu,
y, z)~,~\,
\end{eqnarray}
for all $i=0, 1, 2, ..., m$. So, we only have one independent
$Z_2$ symmetry. And for $i=0$, we always have above equations
for $P_0^z$.

For $l_0 \ge 2$, if $(p_i^z + q_i^z)$ is a multiple of $2^{l_0}$, i. e.,
 $2^{l_0} | (p_i^z+q_i^z)$, we obtain that
\begin{eqnarray}
P_i^z \phi_{\pm \pm \pm +} (x^\mu, y, z) = \phi_{\pm \pm \pm +} (x^\mu, y,
z)~,~\,
\end{eqnarray}
 \begin{eqnarray}
P_i^z \phi_{\pm \pm \pm -} (x^\mu, y, z) = -\phi_{\pm \pm \pm -} (x^\mu,
y, z)~,~\,
\end{eqnarray}
and if  $(p_i^z + q_i^z)$ is not a multiple of $2^{l_0}$,
i. e., $2^{l_0} \not\vert (p_i^z+q_i^z)$
 we obtain that
\begin{eqnarray}
P_i^z \phi_{\pm \pm +\pm} (x^\mu, y, z) = \phi_{\pm \pm +\pm} (x^\mu, y,
z)~,~\,
\end{eqnarray}
 \begin{eqnarray}
P_i^z \phi_{\pm \pm -\pm} (x^\mu, y, z) = -\phi_{\pm \pm -\pm} (x^\mu, y,
z)~.~\,
\end{eqnarray}
 So, we  have two non-equivalent $Z_2$ symmetries along the $z$ direction.

Therefore, we can have at most four non-equivalent $Z_2$ symmetries.
Because we need discrete symmetry to break the bulk gauge symmetry
and supersymmetry, we will concentrate on the scenario with
$k_0 \ge 2$ and $l_0 \ge 2$. To be explicit, we would like to give two
examples:
(I) $n=1$, $m=1$, $4|(p_1^y+q_1^y)$, $4|(p_1^z+q_1^z)$, in this simple
case,
 we can have four local $Z_2$ symmetries. (II) Suppose 
$n=3$, $m=3$, $4|(p_1^y+q_1^y)$, $4|(p_1^z+q_1^z)$, $p_2^y=q_2^y=1$,
$p_2^z=q_2^z=1$, $p_3^y=q_1^y$, $q_3^y=p_1^y$,  $p_3^z=q_1^z$,
and $q_3^z=p_1^z$,
we will have two global $Z_2$ symmetries and
two local $Z_2$ symmetries. The local $Z_2$ symmetry will become
global if $p_1^y=1$ and $q_1^y=3$, or $p_1^z=1$ and $q_1^z=3$.
The two global $Z_2$ symmetries can  be moduloed from the
manifold. In general, we may have global $(Z_2)^{k_0}$ and
$(Z_2)^{l_0}$ symmetries, the extra space orbifold will
be $S^1/(Z_2)^{k_0} \times S^1/(Z_2)^{l_0}$ if we modulo
those global $Z_2$ symmetries.

\subsection{Discrete Symmetry on $M^4\times S^1/Z_2 \times S^1/Z_2$}
In this subsection, 
we would like to discuss the discrete symmetry on the space-time 
$M^4\times S^1/Z_2 \times S^1/Z_2$, where there are some
 3-branes with $Z_2$ symmetry
in the bulk. And we denote two global $Z_2$ symmtries $y \sim -y$ and
$z \sim -z$ as $P^y$ and $P^z$.
For simplicity, we assume that there are only four 4-branes
 which are the boundary branes on $S^1/Z_2 \times S^1/Z_2$, and the
 3-branes are only in the bulk. 

Suppose we have $n$ 3-branes in the bulk, and their coodinates
are $(y_i, z_i)$ where $0~<  ~y_i ~<~ \pi R_1$ and $0~<  ~z_i ~<~ \pi
R_2$.
We denote the local coordinates for the $i$-th 3-brane as
$y_i'\equiv y-y_i$, and $z_i'\equiv z-z_i$. And then,
the equivalent class
for the reflection $Z_2$ symmetry in the $i-th$ 3-brane neighborhood
is $(y_i',~ z_i')~ \sim (-y_i', ~-z_i')$. For that $Z_2$ symmetry,
we define the corresponding
 $Z_2$ operator $P_i$ for $i$= 1,2, ..., n,  whose
eigenvalue is $\pm 1$, i. e., for a generic field or function,
we have
\begin{eqnarray}
P_i \phi(x^{\mu}, -y_i', -z_i') =\pm \phi(x^{\mu}, y_i', z_i')
~.~\,
\end{eqnarray} 

Because if $y_i/(2 \pi R_1)$ or $z_i/(2 \pi R_1)$
 is an irrational number, we will project
out all the KK states, which can not satisfy our requirement. So, we 
assume 
\begin{eqnarray}
y_i = {p^y_i\over q^y_i}  \pi R_1, ~~~~~~~~~~ {\rm for}~ i=1, 2, ..., n  
~,~\,
\end{eqnarray}
\begin{eqnarray}
z_i = {p^z_i\over q^z_i} \pi R_2, ~~~~~~~~~~ {\rm for}~ i=1, 2, ..., m 
~,~\,
\end{eqnarray}
where $p^y_i$ and $q^y_i$ are relative prime positive integers,
and $p^z_i$ and $q^z_i$ are relative prime positive integers.

Assume $L_y$ is the least common multiple for all $p_i^y+q_i^y$, 
and $L_z$ is  the least common multiple for all $p_i^z+q_i^z$, i. e.,
\begin{eqnarray}
L_y \equiv [p_1^y+q_1^y, p_2^y+q_2^y, ..., p_n^y+q_n^y]
~,~\,
\end{eqnarray} 
\begin{eqnarray}
L_z \equiv [p_1^z+q_1^z, p_2^z+q_2^z, ..., p_m^z+q_m^z]
~.~\,
\end{eqnarray} 

By Unique Prime Factorization theorem, we obtain that
\begin{eqnarray}
L_y = 2^{k_0} s_1^{k_1} s_2^{k_{2}} ... s_u^{k_{u}}
~,~\,
\end{eqnarray} 
\begin{eqnarray}
L_z = 2^{l_0} t_1^{l_1} t_2^{l_{2}} ... t_v^{l_{v}}
~,~\,
\end{eqnarray} 
where $2 < s_1 < s_2 <... < s_u$,  $2 < t_1 < t_2 <... < t_v$,
$k_0$ and $l_0$
are the non-negative integers, $s_i$ and $t_j$ are the prime numbers,
and $k_i$ and $l_j$ are the positive integers for $i=1, 2, ..., u$
and $j=1, 2, ..., v$.
And we define the effective physical radii $r_1$ and $r_2$ as
\begin{eqnarray}
r_1 = \left\{ \begin{array}{ll}
2R_1/L_y  & ~~~~{\rm for}~ k_0 \ge 1~, \\ 
R_1/L_y   &~~~~{\rm for}~  k_0=0 ~, \end{array} \right.
\end{eqnarray}  
\begin{eqnarray}
r_2 = \left\{ \begin{array}{ll}
2R_2/L_z  & ~~~~{\rm for}~ l_0 \ge 1~, \\ 
R_2/L_z   &~~~~{\rm for}~  l_0=0 ~. \end{array} \right.
\end{eqnarray}  

Because we require that every fields have zero modes or KK modes, we can
only
have one additional non-equivalent $Z_2$ symmetry for all the 3-branes,
 which is not equivalent
to two global $Z_2$ symmetries $P^y$ and $P^z$. And in this scenario, 
$k_0 \ge 1$, $l_0 \ge 1$; $2^{k_0} | (p_i^y+q_i^y)$
for $i=1, 2, ..., n$, $2^{l_0} | (p_j^z+q_j^z)$
for $j=1, 2, ..., m$. 

Because all the $Z_2$ symmetries for the 3-branes are equivalent,
 we write them as
$P^{y' z'}$.
Denoting the field with ($P^y$, $P^z$, $P^{y' z'}$)=($\pm, \pm, \pm$) by
$\phi_{\pm \pm \pm}$, we obtain the following KK mode expansions
\begin{eqnarray}
  \phi_{+++} (x^\mu, y, z) &=& 
\sum_{n=0}^{\infty} \sum_{m=0}^{\infty}
 \left( \phi_{+++}^{(2n, 2m)}(x^{\mu}) A_{++}^{2n}(y, r_1) A_{++}^{2m}(z,
r_2)
\right.\nonumber\\&&\left.
+\phi^{(2n+1, 2m+1 )}_{+++}(x^\mu) A_{+-}^{2n+1}(y, r_1) A_{+-}^{2m+1}(z,
r_2) \right)
 ~,~\,
\end{eqnarray}
\begin{eqnarray}
  \phi_{++-} (x^\mu, y, z) &=& 
\sum_{n=0}^{\infty} \sum_{m=0}^{\infty}
 \left( \phi_{++-}^{(2n, 2m+1)}(x^{\mu}) A_{++}^{2n}(y, r_1)
A_{+-}^{2m+1}(z, r_2)
\right.\nonumber\\&&\left.
+\phi^{(2n+1, 2m )}_{++-}(x^\mu) A_{+-}^{2n+1}(y, r_1) A_{++}^{2m}(z, r_2)
\right)
 ~,~\,
\end{eqnarray}
\begin{eqnarray}
  \phi_{+-+} (x^\mu, y, z) &=& 
\sum_{n=0}^{\infty} \sum_{m=0}^{\infty}
 \left( \phi_{+-+}^{(2n, 2m+1)}(x^{\mu}) A_{++}^{2n}(y, r_1)
A_{-+}^{2m+1}(z, r_2)
\right.\nonumber\\&&\left.
+\phi^{(2n+1, 2m+2 )}_{+-+}(x^\mu) A_{+-}^{2n+1}(y, r_1) A_{--}^{2m+2}(z,
r_2) \right)
 ~,~\,
\end{eqnarray}
\begin{eqnarray}
  \phi_{+--} (x^\mu, y, z) &=& 
\sum_{n=0}^{\infty} \sum_{m=0}^{\infty}
 \left( \phi_{+--}^{(2n, 2m+2)}(x^{\mu}) A_{++}^{2n}(y, r_1)
A_{--}^{2m+2}(z, r_2)
\right.\nonumber\\&&\left.
+\phi^{(2n+1, 2m+1 )}_{+--}(x^\mu) A_{+-}^{2n+1}(y, r_1) A_{-+}^{2m+1}(z,
r_2) \right)
 ~,~\,
\end{eqnarray}
\begin{eqnarray}
  \phi_{-++} (x^\mu, y, z) &=& 
\sum_{n=0}^{\infty} \sum_{m=0}^{\infty}
 \left( \phi_{-++}^{(2n+1, 2m)}(x^{\mu}) A_{-+}^{2n+1}(y, r_1)
A_{++}^{2m}(z, r_2)
\right.\nonumber\\&&\left.
+\phi^{(2n+2, 2m+1 )}_{-++}(x^\mu) A_{--}^{2n+2}(y, r_1) A_{+-}^{2m+1}(z,
r_2) \right)
 ~,~\,
\end{eqnarray}
\begin{eqnarray}
  \phi_{-+-} (x^\mu, y, z) &=& 
\sum_{n=0}^{\infty} \sum_{m=0}^{\infty}
 \left( \phi_{-+-}^{(2n+1, 2m+1)}(x^{\mu}) A_{-+}^{2n+1}(y, r_1)
A_{+-}^{2m+1}(z, r_2)
\right.\nonumber\\&&\left.
+\phi^{(2n+2, 2m )}_{-+-}(x^\mu) A_{--}^{2n+2}(y, r_1) A_{++}^{2m}(z, r_2)
\right)
 ~,~\,
\end{eqnarray}
\begin{eqnarray}
  \phi_{--+} (x^\mu, y, z) &=& 
\sum_{n=0}^{\infty} \sum_{m=0}^{\infty}
 \left( \phi_{--+}^{(2n+1, 2m+1)}(x^{\mu}) A_{-+}^{2n+1}(y, r_1)
A_{-+}^{2m+1}(z, r_2)
\right.\nonumber\\&&\left.
+\phi^{(2n+2, 2m+2 )}_{--+}(x^\mu) A_{--}^{2n+2}(y, r_1) A_{--}^{2m+2}(z,
r_2) \right)
 ~,~\,
\end{eqnarray}
\begin{eqnarray}
  \phi_{---} (x^\mu, y, z) &=& 
\sum_{n=0}^{\infty} \sum_{m=0}^{\infty}
 \left( \phi_{---}^{(2n+1, 2m+2)}(x^{\mu}) A_{-+}^{2n+1}(y, r_1)
A_{--}^{2m+2}(z, r_2)
\right.\nonumber\\&&\left.
+\phi^{(2n+2, 2m+1 )}_{---}(x^\mu) A_{--}^{2n+2}(y, r_1) A_{-+}^{2m+1}(z,
r_2) \right)
 ~.~\,
\end{eqnarray}
The 4-dimensional fields $\phi^{(n, m)}$ acquire masses 
$\sqrt {n^2/r_1^2+ m^2/r_2^2}$ upon the compactification.
And the
zero modes are contained only in $\phi_{+++}$ fields.
Moreover, because $0 < r_1 \le R_1$, and $0 < r_2 \le R_2$,
 the masses of KK states ($\sqrt {n^2/r_1^2+ m^2/r_2^2}$)
can be set arbitrarily heavy if $L_y$ and $L_z$ are
large enough, i. e.,  we choose suitable $(p_i^y, q_i^y)$ for some $i$,
and $(p_j^z, q_j^z)$ for some $j$.
Therefore, there is no simple relation between the physical size of 
the extra dimensions and the mass scales of KK states.

In short, we can  have three non-equivalent $Z_2$ symmetries where
two are global symmetries.
 To be explicit, we would like to give an example:
$n=1$,  $2|(p_1^y+q_1^y)$, and $2|(p_1^z+q_1^z)$. In this simple
example, the local $Z_2$ symmetry for the 3-brane can become global
if $p_1^y=q_1^y=1$ and $p_1^z=q_1^z=1$, and this global symmetry
can be moduloed, then, the space-time is 
$M^4\times (S^1/Z_2 \times S^1/Z_2)/Z_2$.

\section{GUT Breaking on the Space-Time $M^4\times M^1\times M^1$}
In this section, we would like to discuss the GUT breaking on 
the space-time $M^4\times M^1\times M^1$. 
As we know, the 6-dimensional $N=1$ supersymmetric theory is
chiral, where the gaugino (and gravitino) has positive chirality
and the matters (hypermultiplets) have negative chirality, so, 
it often has anomaly except that we put the
Standard Model fermions on the brane and add a multiplet in
the adjoint representation of gauge group or some other
matter contents in the bulk to cancell the 
gauge anomaly. The 6-dimensional non-supersymmetric GUT models and $N=1$ 
supersymmetric GUT models can be considered as special cases of
$N=2$ supersymmetric GUT models,
 so, we only discuss the
 6-dimensional $N=2$ supersymmetric GUT models.
 $N=2$ supersymmetric 
$SU(5)$, $SU(6)$, $SU(7)$, $SO(10)$, and $SO(12)$ models
on the space-time $M^4\times T^2/(Z_2)^3$ and 
$M^4\times T^2/(Z_2)^4$ have been studied
completely in Ref.~\cite{LIT2},
 and those discussions can be extended to the complete discussions of
 GUT breaking on the space-time $M^4\times M^1\times M^1$
where there are three and four $Z_2$ symmetries.
Because the discussions for GUT breaking are similar, we will not
 make the complete discussions for $SU(M)$ and $SO(2M)$ models
 in this paper. To explain the
idea, we will discuss the
$N=2$ supersymmetric $SU(6)$ and $SO(10)$ models on 
$M^4\times S^1\times S^1$ where there are four $Z_2$ symmetries, and the
$N=2$ supersymmetric $SU(6)$ model with gauge-Higgs unification on 
$M^4\times S^1/Z_2\times S^1/Z_2$ where there are
three $Z_2$ symmetries.

Let us explain the 6-dimensional gauge theory with $N=2$ supersymmetry.
$N=2$ supersymmetric theory in 6-dimension has 16 real supercharges,
corresponding to $N=4$ supersymmetry in 4-dimension. So, only the
vector multiplet can be introduced in the bulk, and we have to
put the Standard Model fermions on the 4-branes, 3-branes or
4-brane intersections.
In terms of the 4-dimensional
$N=1$ language, it contains a vector multiplet $V(A_{\mu}, \lambda_1)$,
and three chiral multiplets $\Sigma_5$, $\Sigma_6$, and $\Phi$. All 
of them are in the adjoint representation of the gauge group. In addition,
the $\Sigma_5$ and $\Sigma_6$ chiral multiplets
contain the gauge fields $A_5$ and $A_6$ in
their lowest components, respectively.

In the Wess-Zumino gauge and 4-dimensional $N=1$ language, the bulk action 
is~\cite{NAHGW}
\begin{eqnarray}
  S &=& \int d^6 x \Biggl\{
  {\rm Tr} \Biggl[ \int d^2\theta \left( \frac{1}{4 k g^2} 
  {\cal W}^\alpha {\cal W}_\alpha + \frac{1}{k g^2} 
  \left( \Phi \partial_5 \Sigma_6 - \Phi \partial_6 \Sigma_5
  - \frac{1}{\sqrt{2}} \Phi 
  [\Sigma_5, \Sigma_6] \right) \right) 
\nonumber\\
&& + {\rm H.C.} \Biggr] 
  + \int d^4\theta \frac{1}{k g^2} {\rm Tr} \Biggl[ 
  \sum_{i=5}^6 \left((\sqrt{2} \partial_i + \Sigma_i^\dagger) e^{-V} 
  (-\sqrt{2} \partial_i + \Sigma_i) e^{V} + 
   \partial_i e^{-V} \partial_i e^{V}\right)
\nonumber\\
  && \qquad \qquad \qquad
  + \Phi^\dagger e^{-V} \Phi e^{V}  \Biggr] \Biggr\} ~.~\,
\label{eq:5daction}
\end{eqnarray}
And the gauge transformation is given by
\begin{eqnarray}
  e^V &\rightarrow& e^\Lambda 
    e^V e^{\Lambda^\dagger}, \\
  \Sigma_i &\rightarrow& e^\Lambda (\Sigma_i - \sqrt{2} \partial_i) 
    e^{-\Lambda}, \\
  \Phi &\rightarrow& e^\Lambda \Phi e^{-\Lambda}~,~\,
\end{eqnarray}
where $i=5, 6$.

From the action, we obtain 
the vector multiplet transformations under the $Z_2$ operators $P^y_i$,
$P^z_j$, $P^{y'z'}$
\begin{eqnarray}
  V(x^{\mu},-y_i', z) &=& P^y_i V(x^{\mu}, y_i', z) (P^y)^{-1}
~,~\,
\end{eqnarray}
\begin{eqnarray}
  \Sigma_5(x^{\mu},-y_i', z) &=& - P^y_i \Sigma_5(x^{\mu}, y_i', z)
(P^y_i)^{-1}
~,~\,
\end{eqnarray}
\begin{eqnarray}
  \Sigma_6(x^{\mu},-y_i', z) &=&  P^y_i \Sigma_6(x^{\mu}, y_i', z)
(P^y_i)^{-1}
~,~\,
\end{eqnarray}
\begin{eqnarray}
  \Phi(x^{\mu},-y_i', z) &=& - P^y_i \Phi(x^{\mu}, y_i', z) (P^y_i)^{-1}
~,~\,
\end{eqnarray}
\begin{eqnarray}
  V(x^{\mu},y, -z_j') &=& P^z_j V(x^{\mu}, y, z_j') (P^z_j)^{-1}
~,~\,
\end{eqnarray}
\begin{eqnarray}
  \Sigma_5(x^{\mu}, y, -z_j') &=&  P^z_j \Sigma_5(x^{\mu}, y, z_j')
(P^z_j)^{-1}
~,~\,
\end{eqnarray}
\begin{eqnarray}
  \Sigma_6(x^{\mu}, y, -z_j') &=&  - P^z_j \Sigma_6(x^{\mu}, y, z_j')
(P^z_j)^{-1}
~,~\,
\end{eqnarray}
\begin{eqnarray}
  \Phi(x^{\mu}, y, -z_j') &=& - P^z_j \Phi(x^{\mu}, y, z_j') (P^z_j)^{-1}
~,~\,
\end{eqnarray}
\begin{eqnarray}
  V(x^{\mu}, -y_i', -z_i') &=& P^{y' z'} V(x^{\mu}, y_i', z_i') (P^{y'
z'})^{-1}
~,~\,
\end{eqnarray}
\begin{eqnarray}
  \Sigma_5(x^{\mu}, -y_i', -z_i') &=& - P^{y' z'} \Sigma_5(x^{\mu}, y_i',
z_i')
 (P^{y' z'})^{-1}
~,~\,
\end{eqnarray}
\begin{eqnarray}
  \Sigma_6(x^{\mu}, -y_i', -z_i') &=&  - P^{y' z'} \Sigma_6(x^{\mu}, y_i',
z_i') 
(P^{y' z'})^{-1}
~,~\,
\end{eqnarray}
\begin{eqnarray}
  \Phi(x^{\mu}, -y_i', -z_i') &=&  P^{y' z'} \Phi(x^{\mu}, y_i', z_i')
(P^{y' z'})^{-1}
~.~\,
\end{eqnarray}

\subsection{$SU(6)$ and $SO(10)$ breaking on $M^4\times S^1\times S^1$}
In this subsection, we would like to discuss the
$SU(6)$ and $SO(10)$ models on $M^4\times S^1\times S^1$.
We require that: (1) 
there are no zero modes for the
chiral multiplets $\Sigma_5$, $\Sigma_6$ and $\Phi$; (2)
 for the zero modes, we only have 4-dimensional
 $N=1$ supersymmetric $SU(3)\times SU(2)\times U(1)^{2}$
model.

For simplicity, we assume that there are four 4-branes, where two along
the
$y$ direction and two along the $z$ direction, $4|(p_1^y+q_1^y)$, and
$4|(p_1^z+q_1^z)$. So, we will have four local $Z_2$ symmetries:
$P_0^y$, $P_1^y$, $P_0^z$ and $P_1^z$.

We will choose the unit matrix representations for $P^y_0$ and $P^z_0$ in
the
adjoint representation of GUT gauge group. So, considering the zero modes,
under $P^y_0$ projection, we can break the 4-dimensional $N=4$
supersymmetry to
$N=2$ supersymmetry with $(V, \Sigma_6)$ forming a vector multiplet and
$(\Sigma_5, \Phi)$ forming a hypermultiplet, and we can break the
4-dimensional $N=2$ supersymmetry to $N=1$ supersymmetry further by
$P^z_0$ projection. 

For a generic bulk field $\phi(x^{\mu}, y, z)$,
we can define four parity operators $P^y_0$, $P^{y}_1$, 
$P^{z}_0$ and $P^{z}_1$, respectively.
Denoting the field with 
($P^y_0$, $P^{y}_1$, $P^{z}_0$, $P^{ z}_1$)=($\pm, \pm, \pm, \pm$) by 
$\phi_{\pm \pm \pm \pm}$, we obtain the KK mode expansions, 
which are those given in Eq.s (61-76). 

(I) $SU(6)$ Model. 
We need to choose the matrix representations 
for parity operators $P^y_0$, $P^y_1$, $P^{z}_0$ and $P^{z}_1$,
 which are expressed in the adjoint representaion of SU(6).
Because $SU(6) \supset SU(5)\times U(1);$ $SU(4)\times SU(2)\times U(1);$
$SU(3)\times SU(3)\times U(1)$, we obtain that, in general, 
$P^{y}_1$ and $P^{z}_1$ just need to be any two different representaions
from
these three representations: ${\rm diag}(+1, +1, +1, +1, +1, -1)$,
${\rm diag}(-1, -1, -1, +1, +1, -1)$, and ${\rm diag}(-1, -1, -1, +1, +1,
+1)$.
So, the matrix representations for $P^y_0$, $P^z_0$, $P^{y}_1$ and
$P^{z}_1$ 
are~\footnote{For $SU(6)$ model and $SO(10)$
model, one can interchange the matrix representations
$P^{y}_1$ and $P^{z}_1$, i. e.,
$P^{y}_1 \longleftrightarrow P^{z}_1$,  and
the discussions are similar.}
\begin{equation}
P^y_0={\rm diag}(+1, +1, +1, +1, +1, +1)
~,~ P^z_0={\rm diag}(+1, +1, +1, +1, +1, +1)~,~\,
\end{equation}
\begin{equation}
P^{y}_1={\rm diag}(+1, +1, +1, +1, +1, -1)
~,~ P^{z}_1={\rm diag}(-1, -1, -1, +1, +1, -1)
 ~,~\,
\end{equation}
or 
\begin{equation}
P^{y}_1={\rm diag}(+1, +1, +1, +1, +1, -1)
~,~ P^{z}_1={\rm diag}(-1, -1, -1, +1, +1, +1)
 ~,~\,
\end{equation}
or
\begin{equation}
P^{y}_1={\rm diag}(-1, -1, -1, +1, +1, +1)
~,~ P^{z}_1={\rm diag}(-1, -1, -1, +1, +1, -1)
 ~.~\,
\end{equation}

And we would like to point out that
\begin{equation}
SU(6)/\{{\rm diag}(+1, +1, +1, +1, +1, -1)\} \approx
SU(5)\times U(1) ~,~\,
\end{equation}
\begin{equation}
SU(6)/\{{\rm diag}(-1, -1, -1, +1, +1, -1)\} \approx
SU(4)\times SU(2) \times U(1) ~,~\,
\end{equation}
\begin{equation}
SU(6)/\{{\rm diag}(-1, -1, -1, +1, +1, +1)\} \approx
SU(3)\times SU(3) \times U(1) ~.~\,
\end{equation}

(II) $SO(10)$ Model. 
We choose the following matrix representations for the parity operators
 $P^y_0$, $P^z_0$, $P^{y}_1$, and $P^{z}_1$,
which are expressed in the adjoint representaion of $SO(10)$
\begin{equation}
P^y_0={\rm diag}(+\sigma_0, +\sigma_0, +\sigma_0, +\sigma_0, +\sigma_0)
~,~\,
\end{equation}
\begin{equation}
P^z_0={\rm diag}(+\sigma_0, +\sigma_0, +\sigma_0, +\sigma_0,
+\sigma_0)~,~\,
\end{equation}
\begin{equation}
P^{y}_1={\rm diag}(\sigma_2, \sigma_2, \sigma_2, \sigma_2, \sigma_2)
~,~\,
\end{equation}
\begin{equation}
P^{z}_1={\rm diag}(-\sigma_0, -\sigma_0, -\sigma_0, +\sigma_0, +\sigma_0)
 ~,~\,
\end{equation}
where $\sigma_0$ is the $2\times 2$ unit matrix and  $\sigma_2$ is the 
Pauli matrix. 

And we would like to point out that
\begin{equation}
 SO(10)/P^{y}_1 \approx  SU(5)\times U(1)~,~\,
\end{equation}
\begin{equation}
 SO(10)/P^{z}_1 \approx  SU(4)\times SU(2)\times SU(2)~.~\,
\end{equation}

Now, we discuss the GUT breaking for $SU(6)$ and $SO(10)$ together.
Assume $G=SU(6)$ or $G=SO(10)$. 
Under $P^{y}_1$ and $P^{z}_1$ parities,
the gauge generators $T^A$, where A=1, 2, ..., 35 for $SU(6)$ and 45 for
$SO(10)$
are separated into four sets: $T^{a, b}$ are the gauge generators for
$SU(3)\times SU(2) \times U(1) \times U(1)$ gauge symmetry, $T^{ a, \hat
b}$,
$T^{\hat a, b}$, and $T^{\hat a, \hat b}$
 are the other broken gauge generators which
belong to $\{G/P^{y}_1 \cap \{{\rm coset}~ G/P^{z}_1\}\}$,
$\{\{{\rm coset}~ G/P^{y}_1\} \cap  G/P^{z}_1\}$, 
and $\{\{{\rm coset}~ G/P^{y}_1\} \cap \{{\rm coset}~ G/P^{z}_1\}\}$,
respectively.
Therefore,
under $P^{y}_0$, $P^{z}_0$, $P^{y}_1$ and $P^{z}_1$, 
the gauge generators transform as
\begin{equation}
P^y_0~T^{A, B}~(P^y_0)^{-1}= T^{A, B} ~,~ P^z_0~T^{A, B}~(P^z_0)^{-1}=
T^{A, B} 
~,~\,
\end{equation}
\begin{equation}
P^{y}_1~T^{a, B}~(P^{y}_1)^{-1}= T^{a, B} ~,~ 
P^{y}_1~T^{\hat a, B}~(P^{y}_1)^{-1}= - T^{\hat a, B}
~,~\,
\end{equation}
\begin{equation}
P^{z}_1~T^{A, b}~(P^{z}_1)^{-1}= T^{A, b} ~,~ 
P^{z}_1~T^{A, \hat b}~(P^{z}_1)^{-1}= - T^{A, \hat b}
~.~\,
\end{equation}

\renewcommand{\arraystretch}{1.4}
\begin{table}[t]
\caption{Parity assignment and masses ($n\ge 0, m \ge 0$) for the
vector multiplet in the $SU(6)$ or $SO(10)$ models on $M^4 \times
S^1\times S^1$.
\label{tab:SUV1}}
\vspace{0.4cm}
\begin{center}
\begin{tabular}{|c|c|c|}
\hline        
$(P^y_0, P^{y}_1, P^z_0, P^{z}_1)$ & field & mass\\ 
\hline
$(+, +, +, +)$ &  $V^{a, b}_{\mu}$ & $\sqrt {(2n)^2/r_1^2+ (2m)^2/r_2^2}$
\\
\hline
$(+, +, +, -)$ &  $V^{a, {\hat b}}_{\mu}$ & $\sqrt {(2n)^2/r_1^2+
(2m+1)^2/r_2^2}$ \\
\hline
$(+, -, +, +)$ &  $V^{{\hat a}, b}_{\mu}$ & $\sqrt {(2n+1)^2/r_1^2+
(2m)^2/r_2^2}$ \\
\hline
$(+,-, +, -)$ &  $V^{\hat{a}, \hat{b}}_{\mu}$ & $\sqrt
{(2n+1)^2/r_1^2+(2m+1)^2/r_2^2}$ \\
\hline
$(-, -, +, +)$ &  $\Sigma_5^{a, b}$ & $\sqrt {(2n+2)^2/r_1^2+
(2m)^2/r_2^2}$ \\
\hline
$(-, -, +, -)$ &  $\Sigma_5^{a, {\hat b}}$ & $\sqrt {(2n+2)^2/r_1^2+
(2m+1)^2/r_2^2}$ \\
\hline
$(-, +, +, +)$ &  $\Sigma_5^{{\hat a}, b}$ & $\sqrt {(2n+1)^2/r_1^2+
(2m)^2/r_2^2}$ \\
\hline
$(-, +, +, -)$ &  $\Sigma_5^{\hat{a}, \hat{b}}$ & $\sqrt {(2n+1)^2/r_1^2+
(2m+1)^2/r_2^2}$ \\
\hline
$(+, +, -, -)$ &  $\Sigma_6^{a, b}$ & $\sqrt {(2n)^2/r_1^2+
(2m+2)^2/r_2^2}$\\
\hline
$(+, +, -, +)$ &  $\Sigma_6^{a, {\hat b}}$ & $\sqrt {(2n)^2/r_1^2+
(2m+1)^2/r_2^2}$\\
\hline
$(+, -, -, -)$ &  $\Sigma_6^{{\hat a}, b}$ & $\sqrt {(2n+1)^2/r_1^2+
(2m+2)^2/r_2^2}$\\
\hline
$(+, -, -,  +)$ &  $\Sigma_6^{\hat{a}, {\hat b}}$ & $\sqrt
{(2n+1)^2/r_1^2+ (2m+1)^2/r_2^2}$ \\
\hline
$(-, -, -, -)$ &  $\Phi^{a, b}$ & $\sqrt {(2n+2)^2/r_1^2+
(2m+2)^2/r_2^2}$\\
\hline
$(-, -, -, +)$ &  $\Phi^{a, {\hat b}}$ & $\sqrt {(2n+2)^2/r_1^2+
(2m+1)^2/r_2^2}$\\
\hline
$(-, +, -, -)$ &  $\Phi^{{\hat a}, b}$ & $\sqrt {(2n+1)^2/r_1^2+
(2m+2)^2/r_2^2}$\\
\hline
$(-, +, -, +)$ &  $\Phi^{\hat{a}, {\hat b}}$ & $\sqrt
{(2n+1)^2/r_1^2+(2m+1)^2/r_2^2}$\\
\hline
\end{tabular}
\end{center}
\end{table}

\renewcommand{\arraystretch}{1.4}
\begin{table}[t]
\caption{For the model $G=SU(6)$ or $G=SO(10)$ on $M^4\times S^1\times
S^1$, the gauge
superfields, the number of 4-dimensional supersymmetry 
and gauge symmetry on the interesections of 4-branes, which
are located at, $(y=0, z=0),$
 $(y=0, z=z_1),$ $(y=y_1, z=0)$, and 
$(y=y_1, z=z_1)$, or on the 4-branes
which are located at 
$y=0$, $z=0$, $y=y_1$, $z=z_1$.
\label{tab:SUV11}}
\vspace{0.4cm}
\begin{center}
\begin{tabular}{|c|c|c|c|}
\hline        
Brane Position & fields & SUSY & Gauge Symmetry\\ 
\hline
$(0, 0) $ &  $V^{A,B}_{\mu}$ & $N=1$ & $G$ \\
\hline
$(0, z_1)$ & $V^{A,b}_{\mu}$, $\Sigma_6^{A, \hat b}$  & N=1 & $G/P^{z}_1$
\\
\hline
$(y_1, 0) $ & $V^{a,B}_{\mu}$, $\Sigma_5^{\hat a, B}$  & N=1 & $G/P^{y}_1
$ \\
\hline
$(y_1, z_1) $ & $V^{a,b}_{\mu}$, $\Sigma_5^{\hat a, b}$,
 $\Sigma_6^{a, \hat b}$, $\Phi^{\hat a, \hat b}$
  & N=1 & $SU(3)\times SU(2)\times U(1) \times U(1)$ \\
\hline
$y=0$ & $V^{A,B}_{\mu}$, $\Sigma_6^{A,B}$  & N=2 & $G$ \\
\hline
$z= 0 $ & $V^{A,B}_{\mu}$, $\Sigma_5^{A,B}$  & N=2 & $G$ \\
\hline
$y=y_1$ & $V^{a,B}_{\mu}$, $\Sigma_5^{\hat a, B}$, $\Sigma_6^{a, B} $,
$\Phi^{\hat a, B}$
  & N=2 & $G/P^{y}_1$ \\
\hline
$z=z_1 $ & $V^{A,b}_{\mu}$, $\Sigma_5^{A,b}$, $\Sigma_6^{A, \hat b} $,
$\Phi^{A, \hat b}$
  & N=2 & $G/P^{z}_1$ \\
\hline
\end{tabular}
\end{center}
\end{table}

The particle spectra are given in Table 2, and the gauge superfields,
the number of 4-dimensional supersymmetry and the gauge group on the
4-branes or 4-brane intersections are given in Table 3.
 For the zero modes, we only have 4-dimensional $N=1$ supersymmetric 
$SU(3)\times SU(2)\times U(1)\times U(1)$ model in the bulk.
And from Table 3, we obtain that including the KK modes,
the 3-brane (the intersection of 4-branes)
 and 4-brane preserve $N=1$ and $N=2$ supersymmetry,
respectively.
The gauge group on the 3-brane can be 
$G$, or $G/P^{z}_1$, or $G/P^{y}_1$, or 
$SU(3)\times SU(2)\times U(1)\times U(1)$.
And the gauge group on the
4-brane can be $G$, or $G/P^{y}_1$ or $G/P^{z}_1$. 
The phenomenology discussions are similar to those in Ref.~\cite{LIT2}.

\subsection{$SU(6)$ Breaking on $M^4\times S^1/Z_2\times S^1/Z_2$}
In this subsection, we would like to discuss the
$SU(6)$ model on $M^4\times S^1/Z_2\times S^1/Z_2$ where there
are two global $Z_2$ symmetries and one local $Z_2$ symmetry. Although
we can not project out all the zero modes for the chiral multiplets
$\Sigma_5$, $\Sigma_6$ and $\Phi$, we require that the extra zero
modes are only from $\Phi$ and form two Higgs doublets, which is called
gauge-Higgs unifications. Our basic requirements
 are: (1) there are no zero modes for the chiral multiplets $\Sigma_5$ 
and $\Sigma_6$; 
(2) considering the zero modes,
there are only one pair of Higgs doublets because if
we had two pairs of Higgs doublets, we may have flavour changing
neutral current problem.  

For simplicity, we assume that there are only four 4-branes, which are the
boundaries for $S^1/Z_2\times S^1/Z_2$, there is only one
3-brane in the bulk, and $2|(p_1^y+q_1^y)$, $2|(p_1^z+q_1^z)$.
So, we will have two global $Z_2$ symmetries
$P^y$ and $P^z$, and one local $Z_2$ symmetry $P^{y'z'}$.
In order to project out all the zero modes of $\Sigma_5$ and $\Sigma_6$,
we would like to choose the matrix representation of $P^{y}$
is equal to that of $P^{z}$.

For a generic bulk field $\phi(x^{\mu}, y, z)$,
we can define three parity operators $P^y$, $P^z$, 
$P^{y'z'}$, respectively.
Denoting the field with 
($P^y$, $P^z$, $P^{y'z'}$)=($\pm, \pm, \pm$) by 
$\phi_{\pm \pm \pm}$, we obtain the KK mode expansions
which are given in Eq.s (102-109). 

There are two scenarios which have gauge-Higgs unifiacation,

(1) We choose the matrix representations for $P^y$, $P^z$
 and $P^{y'z'}$ as following
\begin{equation}
P^y=P^z={\rm diag}(+1, +1, +1, +1, +1, -1)
~,~\,
\end{equation}
\begin{equation}
P^{y'z'}={\rm diag}(-1, -1, -1, +1, +1, +1)~.~\,
\end{equation}

(2) We choose the matrix representations for $P^y$, $P^z$,
 and $P^{y'z'}$ as following
\begin{equation}
P^y=P^z={\rm diag}(-1, -1, -1, +1, +1, -1)
~,~\,
\end{equation}
\begin{equation}
P^{y'z'}={\rm diag}(-1, -1, -1, +1, +1, +1)~.~\,
\end{equation}

Under $P^{y}$ (or $P^z$) and $P^{y'z'}$ parities,
the gauge generators $T^A$, where A=1, 2, ..., 35 for $SU(6)$
are separated into four sets: $T^{a, b}$ are the gauge generators for
$SU(3)\times SU(2) \times U(1) \times U(1)$ gauge symmetry, $T^{ a, \hat
b}$,
$T^{\hat a, b}$, and $T^{\hat a, \hat b}$
 are the other broken gauge generators which
belong to $\{G/P^{y} \cap \{{\rm coset}~ G/P^{y'z'}\}\}$,
$\{\{{\rm coset}~ G/P^{y}\} \cap  G/P^{y'z'}\}$, 
and $\{\{{\rm coset}~ G/P^{y}\} \cap \{{\rm coset}~ G/P^{y'z'}\}\}$,
respectively.
Therefore,
under $P^{y}$, $P^{z}$, and $P^{y'z'}$, 
the gauge generators transform as
\begin{equation}
P^{y}~T^{a, B}~(P^{y})^{-1}= T^{a, B} ~,~ 
P^{y}~T^{\hat a, B}~(P^{y})^{-1}= - T^{\hat a, B}
~,~\,
\end{equation}
\begin{equation}
P^{z}~T^{a, B}~(P^{z})^{-1}= T^{a, B} ~,~ 
P^{z}~T^{\hat a, B}~(P^{z})^{-1}= - T^{\hat a, B}
~,~\,
\end{equation}
\begin{equation}
P^{y'z'}~T^{A, b}~(P^{y'z'})^{-1}= T^{A, b} ~,~ 
P^{y'z'}~T^{A, \hat b}~(P^{y'z'})^{-1}= - T^{A, \hat b}
~.~\,
\end{equation}

\renewcommand{\arraystretch}{1.4}
\begin{table}[t]
\caption{Parity assignment and masses ($n\ge 0, m \ge 0$) for
the vector multiplet in the $SU(6)$ model on 
$M^4\times S^1/Z_2\times S^1/Z_2$.
\label{tab:SUV1}}
\vspace{0.4cm}
\begin{center}
\begin{tabular}{|c|c|c|}
\hline        
$(P^y, P^z, P^{y' z'})$ & field & mass\\ 
\hline
$(+, +, +)$ &  $V^{ab}_{\mu}$,  $\Phi^{\hat{a} b}$ 
 & $\sqrt {(2n)^2/r_1^2+ (2m)^2/r_2^2}$ \\
& &
or $\sqrt {(2n+1)^2/r_1^2+ (2m+1)^2/r_2^2}$\\
\hline
$(+, +, -)$ &  $V^{a \hat{b}}_{\mu}$,  $\Phi^{\hat{a} \hat{b}}$
  & $\sqrt {(2n+1)^2/r_1^2+(2m)^2/r_2^2}$\\
& &
or $\sqrt {(2n)^2/r_1^2+(2m+1)^2/r_2^2}$ \\
\hline
$(+, -, +)$ &  $\Sigma_5^{\hat{a} \hat{b}}$, $\Sigma_6^{a \hat{b}}$
 & $\sqrt {(2n)^2/r_1^2+ (2m+1)^2/r_2^2}$ \\
& & or 
$\sqrt {(2n+1)^2/r_1^2+ (2m+2)^2/r_2^2}$\\
\hline
$(+, -, -)$ & $\Sigma_5^{\hat{a} b}$, $\Sigma_6^{ab}$  & 
$\sqrt {(2n)^2/r_1^2+ (2m+2)^2/r_2^2}$ \\
& & or 
$\sqrt {(2n+1)^2/r_1^2+ (2m+1)^2/r_2^2}$\\
\hline
$(-, +, +)$ &  $\Sigma_5^{a \hat{b}}$, $\Sigma_6^{\hat{a} \hat{b}}$
 & $\sqrt {(2n+1)^2/r_1^2+ (2m)^2/r_2^2}$ \\
& & or 
$\sqrt {(2n+2)^2/r_1^2+ (2m+1)^2/r_2^2}$\\
\hline
$(-, +, -)$ &  $\Sigma_5^{ab}$, $\Sigma_6^{\hat{a} b}$ & 
$\sqrt {(2n+2)^2/r_1^2+ (2m)^2/r_2^2}$ \\
& & or 
$\sqrt {(2n+1)^2/r_1^2+ (2m+1)^2/r_2^2}$\\
\hline
$(-, -, +)$ &  $V^{\hat{a} b}_{\mu}$,  $\Phi^{ab}$
 & $\sqrt {(2n+1)^2/r_1^2+ (2m+1)^2/r_2^2}$ \\
& & or 
$\sqrt {(2n+2)^2/r_1^2+ (2m+2)^2/r_2^2}$\\
\hline
$(-, -, -)$ &  $V^{\hat{a} \hat{b}}_{\mu}$,  $\Phi^{a \hat{b}}$
 & $\sqrt {(2n+2)^2/r_1^2+(2m+1)^2/r_2^2}$\\
& & or $\sqrt {(2n+1)^2/r_1^2+(2m+2)^2/r_2^2}$ \\
\hline
\end{tabular}
\end{center}
\end{table}

We present the particle spectra in Table 4, and the number of
4-dimensional
supersymmetry and gauge symmetry
on the 3-brane, 4-brane intersections and 4-branes in Table 5.
Because the 3-brane is the fixed point under $P^{y' z'}$ symmetry, it
preserves 4-dimensional $N=2$ supersymmetry. And the 4-branes
are the fixed lines under one global symmetry, the 4-brane intersections
and 4-branes preserve 4-dimensional $N=1$ and $N=2$ supersymmetry,
respectively. The phenomenology discussions are also 
similar to those in Ref.~\cite{LIT2}

\renewcommand{\arraystretch}{1.4}
\begin{table}[t]
\caption{For the
 $G=SU(6)$ model with gauge-Higgs unification on 
$M^4\times S^1/Z_2\times S^1/Z_2$.
The gauge superfield $V_{\mu}$, the number of 4-dimensional supersymmetry
and gauge 
symmetry on the 4-brane intersections or 3-brane, which
are located at the $(y=0, z=0),$ $(y=0, z=\pi R_2),$ 
$(y=\pi R_1, z=0),$ $(y=\pi R_1, z=\pi R_2),$ and 
$(y=y_1, z=z_1)$, and on the 4-branes which are located at the fixed lines
$y=0$, $y=\pi R_1$, $z=0$, $z=\pi R_2$.
\label{tab:SUV11}}
\vspace{0.4cm}
\begin{center}
\begin{tabular}{|c|c|c|c|}
\hline        
Brane Position & Fields & SUSY & Gauge Symmetry\\ 
\hline
$(0, 0) $, $(0, \pi R_2)$, $(\pi R_1, 0) $, $(\pi R_1, \pi R_2)$,
&  $V^{a,B}_{\mu}$ & $N=1$ & $G/P^y~ {\rm or}~ G/P^z$ \\
\hline
$(y_1, z_1) $ & $V^{A,b}_{\mu}$ 
  & N=2 & $G/P^{y' z'}$ \\
\hline
$y=0$, $y=\pi R_1$, $z=0$, $z=\pi R_2$ 
& $V^{a,B}_{\mu}$  & N=2 & $G/P^y ~{\rm or}~ G/P^z$ \\
\hline
\end{tabular}
\end{center}
\end{table}

\section{Discrete Symmetry on the Space-Time
$M^4\times$ $A^2$}
As we know, in each point of the 2-dimensional real manifold, there
is an open neighborhood homeomorphic to $R^2$ in the real coordinates
or $C^1$ in the complex coordinates, and its 
rotation group is locally $SO(2)$ or $U(1)$. So, We may 
define the $Z_n$ discrete symmetry on the 2-dimensional manifold in
which $n$ is any positive integer and we can break 
any supersymmetric $SU(M)$ GUT
 models down to the 4-dimensional $N=1$ supersymmetric
$SU(3)\times SU(2) \times U(1)^{M-4}$ models
for the zero modes. One obvious candidate for the extra
space manifold is the disc $D^2$ where there is one 3-brane at
origin and one 4-brane at outer boundary. To be general, we can consider
that 
the extra space manifold is the
annulus where there are two boundary 4-branes.
For simplicity, we denote the annulus as $A^2$. 

Furthermore, we discuss the KK modes expansions on the space-time 
$M^4\times$ (an Segment of $A^2$), and find that
the masses of KK states might be set arbitrarily
heavy if the range of the angle is small enough.

\subsection{Discrete Symmetry on $M^4\times$ $A^2$}
We consider that the extra space manifold is annulus $A^2$.
The convenient coordinates for the annulus
$A^2$ is circular coordinate $(r, \theta)$,
and it is easy to change it to the complex coordinates by $z=r e^{{\rm
i}\theta}$.
We assume that the innner radius of the annulus is $R_1$, and
the outer radius of the annulus is $R_2$. Taking $R_1=0$, we obtain the
Disc $D^2$. Considering $Z_n$ symmetry on $A^2$, we define
\begin{eqnarray}
\omega=e^{{\rm i} {{2 \pi}\over n}} 
~,~\,
\end{eqnarray} 
and we define the generator for $Z_n$ as $\Omega$ which satifies
$\Omega^n=1$.

For a generic bulk multiplet $\Phi$
which fills a representation of the bulk gauge group $G$, we have
\begin{eqnarray}
\Omega \Phi (x^{\mu}, z, \bar z) = \Phi (x^{\mu}, \omega z, \omega^{n-1}
\bar z)
= \eta_{\Phi}
(R_{\Omega})^{l_\Phi} \Phi (x^{\mu}, z, \bar z) 
(R_{\Omega}^{-1})^{m_\Phi}
~,~\,
\end{eqnarray} 
where $\eta_{\Phi} \subset Z_n$, and $R_{\Omega}$ is an element in
the adjoint representation of $G$ which satisfies $R_{\Omega}^n=1$.

For a generic field $\phi (x^{\mu}, z, \bar z)$ with eigenvalue
$\omega^l$ under the operator $\Omega$, we write it as 
$\phi_{\omega^l} (x^{\mu}, z, \bar z)$, i. e.,
\begin{eqnarray}
\Omega \phi_{\omega^l} (x^{\mu}, z, \bar z) = \omega^l \phi_{\omega^l}
(x^{\mu}, z, \bar z)
~.~\,
\end{eqnarray}
And the KK modes expansions for
$\phi_{\omega^l} (x^{\mu}, z, \bar z)$ are
\begin{eqnarray}
\phi_{\omega^l} (x^{\mu}, z, \bar z)=\sum_{j=-\infty}^{\infty}
\sum_{k=1}^{\infty} 
\phi_{\omega^l}^{(jk)} (x^{\mu}) f_{jk}^{\omega^l} (z, \bar z)
~,~\,
\end{eqnarray}
where $l=0, 1, ..., n-1$, and
\begin{eqnarray}
f_{jk}^{\omega^l} (z, \bar z)=\sum_{s=0}^{n-1} \omega^{(n-s)l}
f_{jk}(\omega^s z, \omega^{n-s} \bar z)
~.~\,
\end{eqnarray} 
And the function $f_{jk}(z, \bar z)$ are defined by
\begin{eqnarray}
f_{jk}(z, \bar z) = {\rm J}_{j}(\lambda_{jk} r) e^{{\rm i} j \theta}  
~,~\,
\end{eqnarray} 
or
\begin{eqnarray}
f_{jk}(z, \bar z) = {\rm J}_{j}(\lambda_{jk} |z|) (z/|\bar z|)^j 
~,~\,
\end{eqnarray} 
where ${\rm J}_{j}(\lambda_{jk} r)$ is the first order Bessel function,
and it satisfies the Dirichlet or Neumann boundary condition at $r=R_1$
and $r=R_2$
\begin{equation}
{\rm J}_{j}(\lambda_{jk} r) = 0 ~~{\rm or}~~
{{d {\rm J}_{j}(\lambda_{jk} r)}\over\displaystyle
{d \lambda_{jk} r}} = 0,
~~~~~{\rm for }~~r=R_1 ~~{\rm and}~~ R_2
~.~\,
\end{equation} 
The zero modes are contained only in $\phi_{\omega^0}$ fields, i. e.
$l=0$. 

First, we consider that the extra space manifold is 
a disc $D^2$, i. e., $R_1=0$, so, 
we only have boundary condition at $r=R_2$, and then we
can have all the KK states. There is also one
 fixed point under $Z_n$ symmetry, which is the
center of the disc. One might wonder whether there is an
singularity for $f_{jk}(z, \bar z)$ at origin $r=0$ when $j\not=0$,
however, there is no singularity and we only have zero modes
at origin because 
${\rm J}_0(0)=1$ and ${\rm J}_s(0)=0$ for $s\ge 1$.
By the way, the global $Z_n$ symmetry can
be moduloed from $D^2$, and the corresponding
 orbifold is $D^2/Z_n$.

Second, we consider that the 
extra space manifold is an annulus $A^2$. 
The $Z_n$ symmetry act free on the annulus $A^2$, so, we can obtain
the quotient manifold $A^2/Z_n$ by moduloing the $Z_n$ symmetry. 
We will also have much
less KK states, i. e., a lot of KK states in
the summation might be absent because the
boundary conditions at $r=R_1$ and
$r=R_2$ must be satisfied simultaneously.
 The interesting phenomenology is
that, we might have the scenario in which only a few 
KK states are light and the other KK states are relatively heavy,
so, we may produce the light KK states of the gauge
fields at future colliders.

\subsection{KK modes on $M^4 \times $ (an Segment of $A^2$)}
If the extra space-manifold is an segment of $A^2$, we would
like to discuss the KK modes.
We assume that the innner radius of the annulus is $R_1$, 
the outer radius of the annulus is $R_2$, and the angle $\theta$ is
\begin{eqnarray}
0 \le \theta \le \alpha~ 2\pi  ~,~\,
\end{eqnarray} 
where $0 ~<~ \alpha~ < ~1$.
The KK modes expansion for a generic field $\phi$ is 
\begin{eqnarray}
\phi (x^{\mu}, z, \bar z)=\sum_{j=-\infty}^{\infty} \sum_{k=1}^{\infty} 
\phi^{(jk)} (x^{\mu}) f_{jk} (z, \bar z)
~,~\,
\end{eqnarray}
where
\begin{eqnarray}
f_{jk}(z, \bar z) = {\rm J}_{j/(4\alpha)}(\lambda_{jk} r) 
e^{{\rm i} j \theta/(4\alpha)}  
~,~\,
\end{eqnarray} 
or
\begin{eqnarray}
f_{jk}(z, \bar z) = {\rm J}_{j/(4\alpha)}(\lambda_{jk} |z|) 
(z/|\bar z|)^{j/(4\alpha)}
~.~\,
\end{eqnarray} 
And at $r=R_1$ and $r=R_2$, the function ${\rm
J}_{j/(4\alpha)}(\lambda_{jk} r)$
should satisfy the Dirichlet boundary condition or Neumann condition,
\begin{equation}
{\rm J}_{j}(\lambda_{jk} r) = 0 ~~{\rm or}~~
{{d {\rm J}_{j}(\lambda_{jk} r)}\over\displaystyle
{d \lambda_{jk} r}} = 0
~.~\,
\end{equation} 
In short, if $\alpha$ is very small, then, $j/(4\alpha)$ will be very
large, and
 the KK states may be set arbitrarily heavy, which is similar to that
in Ref.~\cite{KRD}.

We can define the $Z_2$ reflection symmetry on the
 sector of $D^2$ or the segment of $A^2$.
However, we can not define the discrete symmetry $Z_n$ for $n > 2$
on the sector of $D^2$ or  segment of $A^2$,
 so, it is not interesting for us to discuss the
supersymmetric GUT breaking in this case.

\section{GUT breaking on the Space-Time $M^4\times A^2$}
In this section, we would like to discuss the 6-dimensional $N=2$
supersymmetric
 $SU(M)$ GUT models on 
the space-time $M^4\times A^2$ or $M^4\times D^2$. 

In principle, we can break any $SU(M)$
gauge group on the space-time $M^4\times A^2$ or $M^4\times D^2$
  becuase we can choose
$Z_n$ symmetry in which $n$ is very large. As example, we
will discuss the $SU(6)$ models on the space-time
$M^4\times A^2$ or $M^4\times D^2$ with $Z_9$ symmetry, or
one can consider 
the space-time as $M^4\times A^2/Z_9$ or $M^4\times D^2/Z_9$.

The $N=2$ supersymmetry in 6-dimension
 corresponds to $N=4$ supersymmetry in 4-dimension,
thus, only the gauge multiplet can be introduced in the bulk.  This
multiplet can be decomposed under the 4-dimensional
 $N=1$ supersymmetry into a vector
multiplet $V$ and three chiral multiplets $\Sigma$, $\Phi$, and $\Phi^c$
in the adjoint representation, with the fifth and sixth components of the
gauge
field, $A_5$ and $A_6$, contained in the lowest component of $\Sigma$.
The Standard Model fermions are on the boundary 4-brane at $r=R_1$
or $r=R_2$ for the annulus $A^2$, 
and on the 3-brane at origin or on the boundary 4-brane at $r=R_2$ for 
the disc $D^2$.

In the Wess-Zumino gauge and 4-dimensional $N=1$ language, the bulk action 
is~\cite{NAHGW}
\begin{eqnarray}
  S &=& \int d^6 x \Biggl\{
  {\rm Tr} \Biggl[ \int d^2\theta \left( \frac{1}{4 k g^2} 
  {\cal W}^\alpha {\cal W}_\alpha + \frac{1}{k g^2} 
  \left( \Phi^c \partial \Phi   - \frac{1}{\sqrt{2}} \Sigma 
  [\Phi, \Phi^c] \right) \right) + {\rm h.c.} \Biggr] 
\nonumber\\
  && + \int d^4\theta \frac{1}{k g^2} {\rm Tr} \Biggl[ 
  (\sqrt{2} \partial^\dagger + \Sigma^\dagger) e^{-V} 
  (-\sqrt{2} \partial + \Sigma) e^{V}\Biggr]
\nonumber\\
&&+ \int d^4\theta \frac{1}{k g^2} {\rm Tr} \Biggl[
  + \Phi^\dagger e^{-V} \Phi  e^{V}
  + {\Phi^c}^\dagger e^{-V} \Phi^c e^{V} 
\Biggr] \Biggr\}.
\label{eq:t2z6action}
\end{eqnarray}

Notice that we consider $Z_9$ symmetry, then, $\omega=e^{{\rm i}2\pi/9}$.
From above action, we obtain  
the transformations of gauge multiplet under $\Omega$ as
\begin{eqnarray}
  V(\omega z, \omega^8 \bar z) &=& R_{\Omega} V(z, \bar z)
R_{\Omega}^{-1}~,~\,
\end{eqnarray}
\begin{eqnarray}
  \Sigma(\omega z, \omega^8 \bar z) &=& \omega^8 R_{\Omega} 
\Sigma(z, \bar z) R_{\Omega}^{-1}~,~\,
\end{eqnarray}
\begin{eqnarray}
  \Phi(\omega z, \omega^8 \bar z) &=& \omega^{m} R_{\Omega} 
\Phi(z, \bar z) R_{\Omega}^{-1}~,~\,
\end{eqnarray}
\begin{eqnarray}
  \Phi^c(\omega z, \omega^8 \bar z) &=& \omega^{10-m} R_{\Omega} 
\Phi^c(z, \bar z) R_{\Omega}^{-1}~,~\,
\end{eqnarray}
where $0~\le m \le~ 8$, $R_{\Omega}$ is an element in the adjoint
representation of the GUT gauge group and satisfies the equation
$R_{\Omega}^9=1$. To be compatible with our previous discussions in 
section 6, we choose $m=8$, and then,
\begin{eqnarray}
  \Phi(\omega z, \omega^8 \bar z ) &=& \omega^{8} R_{\Omega} \Phi(z, \bar
z)
 R_{\Omega}^{-1}~,~\,
\end{eqnarray}
\begin{eqnarray}
  \Phi^c(\omega z,\omega^8 \bar z ) &=& \omega^{2} R_{\Omega} \Phi^c(z,
\bar z)
 R_{\Omega}^{-1}~.~\,
\end{eqnarray}
 
Now, we would like to discuss the supersymmetric
 $SU(6)$ model. We choose the following matrix 
representations for the $Z_9$ operator $\Omega$, $R_{\Omega}$,
 which are expressed in the adjoint representaion of 
$SU(6)$
\begin{equation}
R_{\Omega}={\rm diag}(\omega^2, \omega^2, \omega^2, \omega^8, \omega^8,
\omega^5)
 ~.~\,
\end{equation}
So, upon the $Z_9$ operator $\Omega$,
 the gauge generators $T^A$ where A=1, 2, ..., 35 for $SU(6)$
are separated into two sets: $T^a$ are the gauge generators for
the $SU(3)\times SU(2)\times U(1)^2$ gauge group, and
 $T^{\hat a}$ are the other broken
gauge generators 
\begin{equation}
R_{\Omega}~T^a~R_{\Omega}^{-1}= T^a ~,~ 
R_{\Omega}~T^{\hat a}~R_{\Omega}^{-1}= - T^{\hat a}
~.~\,
\end{equation}

First, we consider that the extra space manifold is the annulus $A^2$. 
For the zero modes, we have 4-dimensional $N=1$ supersymmetry and
$SU(3)\times SU(2) \times U(1)^2$ gauge symmetry in the bulk and on
the 4-branes at $r=R_1$ and $r=R_2$. Including the
KK states, we will have the 4-dimensional $N=4$ supersymmetry and
 $SU(6)$ gauge symmetry in the bulk, and on
the 4-branes at $r=R_1$ and $r=R_2$. 

Second, we consider that the extra space manifold is
the disc $D^2$. For the zero modes, 
we have 4-dimensional $N=1$ supersymmetry and
$SU(3)\times SU(2) \times U(1)^2$ gauge symmetry in the bulk and on
the 4-brane at $r=R_2$. Including all the
KK states, we will have the 4-dimensional $N=4$ supersymmetry and
 $SU(6)$ gauge symmetry in the bulk, and on
the 4-brane at $r=R_2$. In addition,
 we always have 4-dimensional $N=1$ supersymmetry and
$SU(3)\times SU(2) \times U(1)^2$ gauge symmetry on the 3-brane at origin
in which only the zero modes exist.
 So, if we put the Standard Model fermions on the 3-brane at origin,
 the extra dimensions
can be large and the gauge hierarchy problem can be solved 
for there does not exist the proton decay problem at all.

In order to break the extra $U(1)$ symmetry, we have to
introduce the extra chiral multiplets which are singlets under
the Standard Model gauge symmetry, and use Higgs mechanism.
 And if we considered the chiral model on the observable brane, we will
have to introduce the exotic particles due to the anomaly cancellation.

In short, we can introduce $Z_n$ symmetry to break any supersymmetric
 $SU(M)$ GUT models as long as $n$ is large enough. 
 There are 4-dimensional $N=1$ supersymmetry and
$SU(3)\times SU(2) \times U(1)^{M-4}$ gauge symmetry in the bulk and on
the 4-branes for the zero modes, and on the 3-brane at origin 
in the disc $D^2$ scenario. Including all the KK states,  
we will have the 4-dimensional $N=4$ supersymmetry and
 $SU(M)$ gauge symmetry in the bulk, and on
the 4-branes. In addition, the Standard Model fermions
are on the boundary 4-brane at $r=R_1$
or $r=R_2$ if the extra space manifold is annulus $A^2$, 
and on the 3-brane at origin or on the boundary 4-brane at $r=R_2$ if
the extra space manifold is disc $D^2$.

\section{Discrete Symmetry on the Space-Time $M^4\times T^2$}
In this section, we would like to discuss the 
discrete symmetry on the space-time $M^4\times T^2$.
Because for any point in $T^2$,
there is an open neighborhood which is homeomorphic
to $R^2$, naively, one might think that 
one can introduce any $Z_n$
symmetry on $T^2$. However, this is not true.
In fact, we can prove that
 the only discrete symmetries on the
torus from the rotation group $SO(2)$ or $U(1)$ are $Z_2$,
$Z_3$, $Z_4$, and $Z_6$.

The proof is the following. In the complex coordinates,
the torus $T^2$ can be defined by $C^1$ moduloed the
equivalent classes: $z \sim z+2 \pi R_1$ 
and $z \sim z + 2 \pi R_2  \omega$ where $\omega=e^{{\rm i}\theta}$.
So, we need to discuss the $Z_2$, $Z_3$ and $Z_n$ symmetries
for $n~>~3$ separately.

First, we consider $Z_2$ symmetry, the equivalent class is 
$z \sim -z$, and the two fixed points are $z=0$ and 
$z=\pi R_1 + \pi R_2 \omega$. So, the global $Z_2$
symmetry can be defined on the general torus $T^2$.
And the 3-branes can be located at the fixed points.

Second, we consider $Z_3$ symmetry. We have to 
choose $R_1=R_2=R$. Define $\theta=2\pi/3$, we obtain that
the equivalent class $z \sim z + 2 \pi R  \omega$ 
is equivalent to the equivalent class
$z \sim z + 2 \pi R e^{{\rm i} \pi /3}$. 
And there are
three fixed points: $z=0$, $z=2\pi R e^{{\rm i}\pi/6}/{\sqrt 3}$, and
$z=4 \pi R e^{{\rm i}\pi/6}/{\sqrt 3}$.
The 3-branes can be located at the fixed points.
  
Third, in order to define the $Z_n$ symmetry for $n ~>~ 3$, we have to 
choose $\theta = 2 \pi /n$ and $R_1=R_2=R$. In addition,
$\omega$ should satisfy the following equations
\begin{equation}
2 \pi R \omega = l' 2\pi R + k' 2 \pi R \omega 
 ~,~\,
\end{equation}
\begin{equation}
2 \pi R \omega \omega = l 2\pi R + k 2 \pi R \omega 
 ~,~\,
\end{equation}
where $l$, $k$, $l'$ and $k'$ are integers. 
The first equation
is satisfied by choosing $l'=0$ and $k'=1$. Moreover, from
the second equation, we obtain
\begin{equation}
\omega={{l\pm \sqrt {l^2 + 4 k}}\over\displaystyle 2} ~.~\,
\end{equation}
The complete solutions are $l=0$ and $k=\pm 1$, $l=1$ and
$k=-1$. And then, all the possible $\omega$ are $\pm 1$, $\pm {\rm i}$,
$e^{{\rm i}2\pi/3}$, $e^{{\rm i}2\pi/6}$. Therefore, the discrete 
symmetries $Z_n$ for $n~>~3$ on the
torus from the rotation group $SO(2)$ or $U(1)$  are 
$Z_4$ and $Z_6$. Moreover, for $Z_4$ discrete symmetry, there are
two $Z_4$ fixed points: $z=0$ and $z={\sqrt 2} \pi R e^{{\rm i}\pi/4}$,
two $Z_2$ fixed points: $z=\pi R$ and $z= \pi R e^{{\rm i}\pi/2}$.
For $Z_6$ discrete symmetry, there is one $Z_6$ fixed point
$z=0$, and there are two $Z_3$ fixed points: 
 $z=2\pi R e^{{\rm i}\pi/6}/{\sqrt 3}$ and
$z=4 \pi R e^{{\rm i}\pi/6}/{\sqrt 3}$, three $Z_2$ fixed points:
$z=\sqrt 3 \pi R e^{{\rm i}\pi/6}$, $z=\pi R$ and $z= \pi R e^{{\rm
i}\pi/3}$.
The 3-branes can be located at those fixed points.

For the general $T^2$ defined by the equivalent classes:
 $z \sim z+2 \pi R_1$ 
and $z \sim z + 2 \pi R_2 e^{i\theta}$, the KK modes
expansion for a generic field $\phi$ is
\begin{eqnarray}
\phi(x^{\mu}, z, \bar z)= \sum_{j=-\infty}^{+\infty}
\sum_{k=-\infty}^{+\infty} \phi^{(jk)}(x^{\mu})
f_{jk}(z, \bar z)~,~\,
\end{eqnarray} 
where
\begin{eqnarray}
f_{jk}(z, \bar z) = {\rm exp}\{{\rm i}[(a-{\rm i}b)z+(a+{\rm i}b) \bar
z]\}
~,~\,
\end{eqnarray} 
where
\begin{eqnarray}
a={j\over {2R_1}}~,~\,
\end{eqnarray}
\begin{eqnarray}
b={1\over {\sin\theta}} ({k\over {2R_2}}
 - {j\over {2R_1}} \cos\theta) ~.~\,
\end{eqnarray}
The mass for $\phi^{jk}$ is
\begin{eqnarray}
M_{\phi^{jk}}^2 = {1\over {\sin\theta^2}}
[{{j^2}\over {4R_1^2}} + {{k^2}\over {4R_2^2}}-
{{2jk}\over {4R_1 R_2}} \cos\theta] ~.~\,
\end{eqnarray}
The zero modes are contained in $\phi^{(00)}$ sector, i. e., $j=0$ and
$k=0$.

The fundamental group for the torus is $Z\bigoplus Z$, so, one
might think we can break the gauge symmetry  by
Wilson line in the mean time. The key question is how to
define the suitable KK modes' expansions for the bulk
fields in the Wilson line approach, because we require that, the fields
without
zero modes, should have KK modes' excitations.

\section{GUT Breaking on the Space-Time $M^4\times T^2$}
We would like to consider the 6-dimensional $N=2$ supersymmetric
$SU(5)$ model on $M^4\times T^2$ with $Z_6$ symmetry.

Notice that we consider $Z_6$ symmetry, we define  $\omega=e^{{\rm
i}2\pi/6}$.
From the action in Eq. (162), we obtain  
the transformations of gauge multiplet under $\Omega$ as
\begin{eqnarray}
  V(\omega z, \omega^5 \bar z) &=& R_{\Omega} V(z, \bar z)
R_{\Omega}^{-1}~,~\,
\end{eqnarray}
\begin{eqnarray}
  \Sigma(\omega z, \omega^5 \bar z) &=& \omega^5 R_{\Omega} 
\Sigma(z, \bar z) R_{\Omega}^{-1}~,~\,
\end{eqnarray}
\begin{eqnarray}
  \Phi(\omega z, \omega^5 \bar z) &=& \omega^{5} R_{\Omega} 
\Phi(z, \bar z) R_{\Omega}^{-1}~,~\,
\end{eqnarray}
\begin{eqnarray}
  \Phi^c(\omega z, \omega^5 \bar z) &=& \omega^2 R_{\Omega} 
\Phi^c(z, \bar z) R_{\Omega}^{-1}~,~\,
\end{eqnarray}
where $R_{\Omega}$ is an element in the adjoint
representation of $SU(5)$ and satisfies the equation
$R_{\Omega}^6=1$. 
 
 We choose the following matrix 
representations for the $Z_6$ operator $\Omega$, $R_{\Omega}$,
 which are expressed in the adjoint representaion of 
$SU(5)$
\begin{equation}
R_{\Omega}={\rm diag}(1, 1, 1, -1, -1)
 ~.~\,
\end{equation}
So, upon the $Z_6$ operator $\Omega$,
 the gauge generators $T^A$ where A=1, 2, ..., 24 for $SU(5)$
are separated into two sets: $T^a$ are the gauge generators for
the Standard Model gauge group, and
 $T^{\hat a}$ are the other broken
gauge generators 
\begin{equation}
R_{\Omega}~T^a~R_{\Omega}^{-1}= T^a ~,~ 
R_{\Omega}~T^{\hat a}~R_{\Omega}^{-1}= - T^{\hat a}
~.~\,
\end{equation}
In addition, the representaion of the generator for $Z_2$ symmetry is 
\begin{equation}
R_{\Omega}^3={\rm diag}(1, 1, 1, -1, -1)
 ~,~\,
\end{equation}
and the representaion of the generator for $Z_3$ symmetry is 
\begin{equation}
R_{\Omega}^2={\rm diag}(1, 1, 1, 1, 1)
 ~.~\,
\end{equation}
Therefore, the gauge symmetries on the 3-branes at $Z_2$ fixed points 
and on the 3-branes at $Z_3$ fixed points are
\begin{equation}
SU(5)/R_{\Omega}^3 \approx SU(3)\times SU(2) \times U(1)
 ~,~\,
\end{equation}
\begin{equation}
SU(5)/R_{\Omega}^2 \approx SU(5)
 ~.~\,
\end{equation}

\renewcommand{\arraystretch}{1.4}
\begin{table}[t]
\caption{For the
 $G=SU(5)$ model on 
$M^4\times T^2$ with $Z_6$ symmetry.
The gauge multiplet, the number of 4-dimensional supersymmetry and gauge 
symmetry on the 3-branes, which
are located at 
 the $Z_6$ fixed point
$z=0$; the $Z_3$ fixed points: 
 $z=2\pi R e^{{\rm i}\pi/6}/{\sqrt 3}$, and
$z=4 \pi R e^{{\rm i}\pi/6}/{\sqrt 3}$; and the $Z_2$ fixed points:
$z=\sqrt 3 \pi R e^{{\rm i}\pi/6}$, $z=\pi R$, and $z= \pi R e^{{\rm
i}\pi/3}$.
\label{tab:SUV11}}
\vspace{0.4cm}
\begin{center}
\begin{tabular}{|c|c|c|c|}
\hline        
Brane Position & Fields & SUSY & Gauge Symmetry\\ 
\hline
$z=0$ & $V^a$ & $N=1$ & $SU(3)\times SU(2) \times U(1)$ \\
\hline
$z=2\pi R e^{{\rm i}\pi/6}/{\sqrt 3}$, 
& $V^A$, $\Sigma^a$, $\Phi^a$, $(\Phi^c)^{\hat a}$ & $N=1$
& $SU(5)$ \\
$z=4 \pi R e^{{\rm i}\pi/6}/{\sqrt 3}$ & & & \\
\hline
$z=\sqrt 3 \pi R e^{{\rm i}\pi/6}$, 
& $V^a$, $\Sigma^A$, $\Phi^A$, $(\Phi^c)^{A}$ & $N=4$
& $SU(3)\times SU(2)\times U(1)$ \\
$z=\pi R$, $z=\pi R e^{{\rm i}\pi/3}$ & & & \\
\hline
\end{tabular}
\end{center}
\end{table}

The gauge multiplet, the number of 4-dimensional supersymmetry and gauge 
symmetry on the 3-branes are given in Table 6.
In short, 
we have 4-dimensional $N=1$ supersymmetry and
the Standard Model gauge symmetry in the bulk for the
zero modes, and on the 3-brane
at $Z_6$ fixed point for all the modes. Including the
KK states, we will have the 4-dimensional $N=4$ supersymmetry and
 $SU(5)$ gauge symmetry in the bulk, the
4-dimensional $N=1$ supersymmetry and $SU(5)$ gauge symmetry on
the 3-branes at $Z_3$ fixed points, and the 4-dimensional
$N=4$ supersymmetry and $SU(3)\times SU(2) \times U(1)$ gauge 
symmetry on the 3-branes at $Z_2$ fixed points.
The Standard Model fermions and Higgs fields can be on any 3-brane
at one of the fixed points. In particular, if we put the Standard Model
fermions and Higgs fields on the 3-brane at $Z_6$ fixed point, the
extra dimensions can be large and the gauge hierarchy problem can be
solved
because there is no proton decay problem at all.

\section{Discussion and Conclusion}

With the ansatz that there exist local or global discrete symmetries in
the 
special branes' neighborhoods, we discuss the general reflection 
$Z_2$ symmetries on
the space-time
 $M^4\times M^1$ and $M^4\times M^1 \times M^1$. 
We obtain that we can have at most two $Z_2$ symmetries on $M^4\times M^1$
and four $Z_2$ symmetries on  $M^4\times M^1 \times M^1$.
As representatives, we discuss the $N=1$ supersymmetric $SU(5)$ model
on the space-time $M^4\times M^1$ where the Standard Model fermions
 can be in the bulk or on the 3-brane, the $N=2$ supersymmetric $SU(6)$ 
and $SO(10)$ models on the space-time $M^4\times S^1 \times S^1$
and the $N=2$ supersymmetric $SU(6)$ model with gauge-Higgs
unification on the space-time $M^4\times S^1/Z_2 \times S^1/Z_2$,
where the Standard Model fermions must be on the 4-brane, or 3-brane,
 or 4-brane intersection,.
For the zero modes, we have 4-dimensional $N=1$
supersymmetry and $SU(3)\times SU(2) \times U(1)^{n-3}$ gauge symmetry
in which $n$ is the rank of GUT gauge group. The gauge symmetry
and supersymmetry may be broken on the 3-branes, or 4-branes, or 4-brane
intersections. In particular, in those models, the 
extra dimensions can be large and the masses of KK states can
be set arbitrarily heavy.

In addition, we discuss the discrete $Z_n$ symmetry on the
space-time $M^4\times A^2$ and $M^4\times D^2$ in which $n$
is any positive integer. 
 In this kind of scenarios, we can break any $SU(M)$ gauge symmetry
 for $M~\ge~5$
 down to the $SU(3)\times SU(2)\times U(1)^{M-4}$ gauge symmetry
by introducing the global $Z_n$ symmetry as long as $n$ is large enough.
In general, considering the 6-dimensional $N=2$ supersymmetry,
 we have 4-dimensional $N=1$ supersymmetry and
$SU(3)\times SU(2) \times U(1)^{M-4}$ gauge symmetry in the bulk and on
the 4-branes for the zero modes, and on the 3-brane at origin
where only the zero modes exist in 
the disc $D^2$ scenario.
Including all the KK states,  
we will have 4-dimensional $N=4$ supersymmetry and
 $SU(M)$ gauge symmetry in the bulk, and on
the 4-branes. The Standard Model fermions should be on the boundary
4-brane or 3-brane at origin. By the way,
 if we put the Standard Model fermions on the 3-brane at origin,
 the extra dimensions
can be large and the gauge hierarchy problem can be solved 
for there does not exist the proton decay problem at all.
Moreover, if the extra space manifold is annulus $A^2$, 
for suitable choices of the inner radius and outer radius,
we might construct the models where only a few KK states are light
and the other KK states are relatively heavy due to the boundary
conditions on the inner and outer boundaries, so, we might
produce the light KK states of gauge fields at future
colliders, which is very interesting in collider physics.

And if the extra space manifold is 
an sector of $D^2$ or an segment of $A^2$, we point out that
 the masses of KK states can be set arbitrarily heavy
if the range of angle is small enough.

Furthermore, we discuss the complete global discrete symmetry 
on the space-time $M^4\times T^2$.
We prove that the possible global discrete symmetries on the torus is
$Z_2$,
$Z_3$, $Z_4$, and $Z_6$. We also discuss the 6-dimensional $N=2$
supersymmetric $SU(5)$ models on the space-time
$M^4\times T^2$ with $Z_6$ symmetry. 
There are 4-dimensional $N=1$ supersymmetry and
the Standard Model gauge symmetry in the bulk for the
zero modes, and on the 3-brane
at $Z_6$ fixed point for all the modes. Including the
KK states, we will have the 4-dimensional $N=4$ supersymmetry and
 $SU(5)$ gauge symmetry in the bulk, the
4-dimensional $N=1$ supersymmetry and $SU(5)$ gauge symmetry on
the 3-branes at $Z_3$ fixed points, and the 4-dimensional
$N=4$ supersymmetry and $SU(3)\times SU(2) \times U(1)$ gauge 
symmetry on the 3-branes at $Z_2$ fixed points.
The Standard Model fermions and Higgs fields can be on any 3-brane
at one of the fixed points. In particular, if we put the Standard Model
fermions and Higgs fields on the 3-brane at $Z_6$ fixed point, the
extra dimensions can be large and the gauge hierarchy problem can be
solved
because there is no proton decay problem at all.

The phenomenology in those scenarios deserve further study.

\section*{Acknowledgments}
This research was supported in part by the U.S.~Department of Energy under
 Grant No.~DOE-EY-76-02-3071.

\end{document}